\input lanlmac
\input epsf.tex
\overfullrule=0pt

\newcount\figno
\figno=0
\def\fig#1#2#3{
\par\begingroup\parindent=0pt\leftskip=1cm\rightskip=1cm\parindent=0pt
\baselineskip=11pt
\global\advance\figno by 1
\midinsert
\epsfxsize=#3
\centerline{\epsfbox{#2}}
\vskip 12pt
{\bf Fig.\ \the\figno:} #1\par
\endinsert\endgroup\par
}
\def\figlabel#1{\xdef#1{\the\figno}%
\writedef{#1\leftbracket \the\figno}%
}
\def\omit#1{}

\def\pre#1{{\tt
arXiv:#1}}%use this to give preprint # in refs

\def\Rc{{\check R}}

\def\qed{\nobreak\hfill\vbox{\hrule height.4pt%
\hbox{\vrule width.4pt height3pt \kern3pt\vrule width.4pt}\hrule height.4pt}\medskip\goodbreak}
% References
\lref\Rob{D.~Robbins,
{\sl Symmetry Classes of Alternating Sign Matrices},
\pre{math.CO/0008045}.}
\lref\RSa{A.~V.~Razumov and Yu.~G.~Stroganov,
{\sl Spin chain and combinatorics},
{\it J. Phys.} A 34 (2001) 3185--3190,
\pre{cond-mat/0012141}.}
\lref\RS{A.V. Razumov and Yu.G. Stroganov, 
{\sl Combinatorial nature
of ground state vector of $O(1)$ loop model},
{\it Theor. Math. Phys.} 
138 (2004) 333--337; {\it Teor. Mat. Fiz.} 138 (2004) 395--400, \pre{math.CO/0104216}.}
\lref\dG{J. de Gier, {\sl Loops, matchings and alternating-sign matrices},
{\it Discr. Math.} 298 (2005) 365--388,
\pre{math.CO/0211285}.}
\lref\BdGN{M.T. Batchelor, J. de Gier and B. Nienhuis,
{\sl The quantum symmetric XXZ chain at $\Delta=-1/2$, alternating sign matrices and 
plane partitions},
{\it J. Phys.} A 34 (2001) L265--L270,
\pre{cond-mat/0101385}.}
\lref\IZER{A. Izergin, {\sl Partition function of the six-vertex
model in a finite volume}, {\it Sov. Phys. Dokl.} 32 (1987) 878--879.}
\lref\KOR{V. Korepin, {\sl Calculation of norms of Bethe wave functions},
{\it Comm. Math. Phys.} 86 (1982) 391--418.}
\lref\FR{I.B.~Frenkel and N.~Reshetikhin, {\sl Quantum affine Algebras and Holonomic 
Difference Equations},
{\it Commun. Math. Phys.} 146 (1992) 1--60.}
\lref\DFZJ{P.~Di Francesco and P.~Zinn-Justin, {\sl Around the Razumov--Stroganov conjecture:
proof of a multi-parameter sum rule}, {\it Elec. J. Comb.} 12 (1) (2005) R6,
\pre{math-ph/0410061}.}
\lref\OPRS{A.V. Razumov and Yu.G. Stroganov, {\sl O(1) loop model with different boundary con- 
ditions and symmetry classes of alternating-sign matrices}, {\it Theor. Math. Phys.} 142 
(2005) 237--243; {\it Teor. Mat. Fiz.} 142 (2005) 284--292, \pre{cond-mat/0108103}.}
\lref\DFOP{P. Di Francesco, {\sl Inhomogenous loop models with open boundaries},
{\it J. Phys.} A 38 (2005) 6091--6120, \pre{math-ph/0504032}.}
\lref\Pas{V.~Pasquier, {\sl Quantum incompressibility and Razumov Stroganov type conjectures},
{\it Ann. Henri Poincar\'e} 7 (2006) 397--421,
\pre{cond-mat/0506075}.}
\lref\DFZJc{P.~Di Francesco and P.~Zinn-Justin, {\sl Quantum Knizhnik--Zamolodchikov equation,
generalized Razumov--Stroganov sum rules and extended Joseph polynomials}, 
{\it J. Phys.} A 38
(2006) L815--L822, \pre{math-ph/0508059}.}
\lref\DF{P.~Di Francesco, {\sl A refined Razumov--Stroganov conjecture},
{\it J. Stat. Mech.} (2004) P08009, \pre{cond-mat/0407477}.}
\lref\STEM{J. Stembridge, {\sl Nonintersecting paths, Pfaffians and plane partitions},
{\it Adv. in Math.} 83 (1990) 96--131.}
\lref\Bre{D. Bressoud, {\sl Proofs and Confirmations: The Story of
the Alternating Sign Matrix Conjecture}, Cambridge Univ. Pr., 1999.}
\lref\DFZJZ{P. Di Francesco, P. Zinn-Justin and J.-B. Zuber,
{\sl Sum rules for the ground states of the $O(1)$ loop model
on a cylinder}, 
{\it J. Stat. Mech.} (2006) P08011, \pre{math-ph/0603009}.}
\lref\ORBI{P. Di Francesco and P. Zinn-Justin,
{\sl From Orbital Varieties to Alternating Sign Matrices},
extended abstract for FPSAC'06, \pre{math-ph/0512047}.}
\lref\Ku{G. Kuperberg, {\sl Symmetry classes of alternating-sign
matrices under one roof}, 
{\it Ann. of Math.} 156 (3) (2002) 835--866,
\pre{math.CO/0008184}.}
\lref\ZJDF{ P. Zinn-Justin and P.~Di Francesco, {\sl Quantum Knizhnik--Zamolodchikov, Totally Symmetric 
Self-Complementary Plane Partitions and Alternating Sign Matrices}, 
accepted for publication in {\it Theor. Math. Phys.},
\pre{math-ph/0703015}.}
\lref\DFBQKZ{P.~Di Francesco, {\sl Boundary $q$KZ equation and generalized 
Razumov-Stroganov sum rules for open IRF models},
{\it J. Stat. Mech.} (2005) P09004, \pre{math-ph/0509011}. }
\lref\DFTS{P.~Di Francesco, {\sl Totally Symmetric Self Complementary Plane Partitions
and the quantum Knizhnik--Zamolodchikov equation: a conjecture},
{\it J. Stat. Mech.} (2006) P09008,  \pre{cond-mat/0607499}.}
\lref\DFVS{P.~Di Francesco, {\sl Open boundary quantum Knizhnik--Zamolodchikov equation
and the weighted enumeration of plane partitions with symmetries}, 
{\it J. Stat. Mech.} (2007) P01024,
\pre{math-ph/0611012}.} 
\lref\RStr{A.V. Razumov and Yu.G. Stroganov, 
{\sl $O(1)$ loop model with different boundary conditions and symmetry classes 
of alternating-sign matrices},
{\it Theor. Math. Phys.} 
{\bf 142} (2005) 237--243; {\it Teor. Mat. Fiz.} 142 (2005) 284--292,
\pre{cond-mat/0108103}.}
\lref\PRdG{P. Pearce, V. Rittenberg and J. de Gier,
{\sl Critical Q=1 Potts model and Temperley--Lieb stochastic processes}, 
\pre{cond-mat/0108051}; P. Pearce, V. Rittenberg, 
J. de Gier and B.~Nienhuis, 
{\sl Temperley--Lieb Stochastic Processes},
{\it J. Phys.} A 35 (2002) L661--L668,
\pre{math-ph/0209017}.}
\lref\MRR{W. Mills, D. Robbins and H. Rumsey, {\sl Enumeration of a symmetry class of 
plane partitions}, {\it Discrete Math.} 67 (1987) 43--55.}
\lref\LGV{B. Lindstr\"om, {\it On the vector representations of
induced matroids}, {\it Bull. London Math. Soc.} 5 (1973)
85--90\semi
I. M. Gessel and X. Viennot, {\it Binomial determinants, paths and
hook formulae}, {\it Adv. in Math.} 58 (1985) 300--321.}
\lref\DFZJb{P. Di Francesco and P. Zinn-Justin,
{\sl Inhomogeneous model of crossing loops and multidegrees of some algebraic varieties},
{\it Commun. Math. Phys.} 262 (2006) 459--487,
\pre{math-ph/0412031}.}
\lref\KZJ{A. Knutson and P. Zinn-Justin, {\sl A scheme related to the Brauer loop model}, 
{\it Adv. in Math.} 214 (2007) 40--77,
\pre{math.AG/0503224}.}
\lref\KZJb{A. Knutson and P. Zinn-Justin, {\sl Brauer loop scheme and orbital varieties}.}
\lref\KM{A.~Knutson and E.~Miller, {\sl Gr\"obner geometry of Schubert polynomials},
{\it Annals of Mathematics} (2003), \pre{math.AG/0110058}.}
\lref\CK{M. Ciucu and K. Krattenthaler, {\sl Plane partitions: $5{1\over 2}$ symmetry classes},
in ``Combinatorial Methods in Representation Theory'', 
(M. Kashiwara, K. Koike, S. Okada, I. Terada, and
H. Yamada, Eds.), Advanced Studies in Pure Mathematics 
{\bf 28} (2000) 81--103, \pre{math.CO/9808018}.}
\lref\KK{C. Krattenthaler,
{\sl Advanced determinant calculus}, S\'eminaire Lotharingien Combin. ("The Andrews Festschrift")
42 (1999) B42q.}
\lref\CEKZ{M. Ciucu, T. Eisenk\"olbl, C. Krattenthaler
and D. Zare, {\sl Enumeration of lozenge tilings of hexagons with a central triangular hole},
{\it J. Combin. Theory} Ser. A 95 (2001) 251--334, \pre{math.CO/9912053};
C. Krattenthaler, {\sl Descending plane partitions and rhombus tilings of a
hexagon with triangular hole}, {\it Europ. J. Combin.} 27 (2006) 1138--1146, \pre{math.CO/0310188}.}
\lref\KAPA{M. Kasatani and V. Pasquier, {\sl On polynomials interpolating between the stationary 
state of a O(n) model and a Q.H.E. ground state}, \pre{cond-mat/0608160}.}
\lref\Zeil{D.~Zeilberger, 
{\sl Proof of a Conjecture of Philippe Di Francesco and Paul Zinn-Justin 
related to the $q$KZ equations and to Dave Robbins' Two Favorite Combinatorial Objects}
in %
{\tt http://www.math.rutgers.edu/$\sim$zeilberg/pj.html}.}
\lref\Smi{F.A.~Smirnov,
{\sl A general formula for soliton form factors in the quantum sine-Gordon
model}, {\it J. Phys.} A 19 (1986) L575--L578.}
\lref\SKLY{E. Sklyanin, {\sl Boundary conditions for integrable quantum systems},
{\it J. Phys.} A 21 (1988) 2375--2389.}
\lref\JAPOP{M. Jimbo, R. Kedem, H. Konno, T. Miwa and R. Weston, 
{\sl Difference Equations in Spin Chains with a Boundary}, 
{\it Nucl. Phys.} B 448 (1995) 429--456,
\pre{hep-th/9502060}.}
\lref\TLAMEAN{P. Di Francesco, O. Golinelli and E. Guitter, 
{\sl Meanders and the Temperley-Lieb algebra},
{\it Comm. Math. Phys.} 186 (1997) 1--59, \pre{hep-th/9602025}.}
%
%some qKZ integrals papers
%
\lref\JMbook{M.~Jimbo and T.~Miwa,
{\it Algebraic Analysis of Solvable Lattice Models\/}
(AMS, Providence, 1995).}
\lref\JM{M.~Jimbo and T.~Miwa,
{\sl Quantum KZ equation with $|q|=1$ and correlation functions of the XXZ model in the gapless regime},
{\it J. Phys.} A (1996) 2923--2958.}
\lref\TV{V.~Tarasov and A.~Varchenko,
{\sl Geometry of $q$-Hypergeometric functions, Quantum Affine Algebras and Elliptic Quantum Groups},
{\it Asterisque} 246 (1997) 1--135, \pre{math.QA/9703044}\semi
Geometry of $q$-Hypergeometric Functions as a Bridge between Yangians and Quantum Affine Algebras,
{\it Invent. Math.} 128 (1997) 501--588, \pre{math.QA/9604011}.}
\lref\MRR{W.H.~Mills, D.P.~Robbins and H.~Rumsey, {\it J. Combin. Theory} Ser. A 34 (1983),
340--359.}
\Title{}
{\vbox{
\centerline{Quantum Knizhnik--Zamolodchikov Equation: }
\medskip
\centerline{Reflecting boundary conditions and Combinatorics} 
}}
\bigskip\bigskip
\centerline{P.~Di~Francesco and P.~Zinn-Justin} 
\medskip
%\centerline{\it  Service de Physique Th\'eorique de Saclay,}
%\centerline{\it CEA/DSM/SPhT, URA 2306 du CNRS,}
%\centerline{\it F-91191 Gif sur Yvette Cedex, France}
\bigskip
\vskip0.5cm
%abstract
\noindent We consider the level 1 solution of quantum Knizhnik--Zamolodchikov equation with reflecting
boundary conditions
which is relevant to the Temperley--Lieb model of loops on a strip. By use of integral formulae we prove
conjectures relating it to the weighted enumeration of Cyclically Symmetric Transpose Complement Plane
Partitions and related combinatorial objects.

\bigskip

%AMS Subject Classification (2000): Primary 05A19; Secondary 82B20
%\draft
\Date{09/2007}
%
%%%%%%%%%%%%%%%%%%%%%%%%%%%%%%%%%%%%%%%%%%%%%%%%%%%%%%%%%%%%%%%%%%%%%
%
\newsec{Introduction}
Since the papers \refs{\RSa,\BdGN}, there has been a great deal of work on the combinatorial
interpretation of quantum integrable models at special points of their parameter space.
The original observation is that the numbers of Alternating Sign Matrices (ASM) and Plane Partitions (PP)
in various symmetry classes appear naturally in the ground state entries 
of the Temperley--Lieb $O(\tau=1)$ model of (non-crossings) loops with various boundary conditions (and related models).
The appearance of ASM numbers was developed further and to some extent explained by the
Razumov--Stroganov conjecture \RS\ and variants \refs{\OPRS,\PRdG} 
interpreting each ground state entry as a number
of certain subsets of ASM. The role of plane partitions remained more obscure until
the recent work
\refs{\DFTS,\DFVS} which showed that the enumeration of symmetry classes of PP also occurs naturally on condition that one consider a slightly more general problem, namely
the quantum Knizhnik--Zamolodchikov equation ($q$KZ), first introduced in this context in
\DFZJc, and in which the parameter $\tau$ is now free.
This provided a (conjectural) bridge between enumerations of symmetry classes of ASM and PP,
which is a fascinating topic of enumerative combinatorics in itself.

The present work is concerned more specifically with the case of the Temperley--Lieb loop model
(and its $q$KZ generalization) defined on a strip with reflecting boundary conditions
(the case of periodic boundary conditions was treated similarly in \ZJDF).
The corresponding ASM were discovered in \refs{\OPRS,\dG}:
they are Vertically Symmetric Alternating Sign Matrices (VSASM) of size $(2n+1)\times(2n+1)$ 
in even strip size $N=2n$, and modified VSASM of size $(2n+1)\times(2n+3)$ in odd strip size $N=2n+1$.
As to the PP, they were discussed in \DFVS: they are 
Cyclically Symmetric Transpose Complement Plane Partitions (CSTCPP) \MRR\ in odd strip size,
and certain modified CSTCPP (referred to as CSTCPP$^\triangle$ in the following) in even strip size.
The conjectures of \DFVS\ concerning the $\tau$-enumeration of these plane partitions 
are the main subject of this work. In Sect.~2 we shall review the basics of integrable loop models
based on the Temperley--Lieb algebra; in Sect.~3 we shall discuss the related $q$KZ equation, 
and review the conjectures of \DFVS;
in Sect.~4 we introduce the main technical tool, that is certain explicit integrals solving $q$KZ;
and finally in Sect.~5 and 6 we prove the conjectures of \DFVS, considering separately even and odd cases.

\newsec{Loop model with reflecting boundary conditions and link patterns}
\subsec{Dense Loop model on a strip}
We consider the version with reflecting boundaries\foot{These boundary conditions are sometimes called ``open'', 
in reference to the equivalent open XXZ spin chain, or ``closed'', due to the way the loops close at the boundaries
of the strip.}
of the inhomogeneous $O(1)$ non-crossing 
loop model \DFOP. The model is defined on a semi-infinite strip of width $N$ (even or odd) of square lattice, 
with centers of the lower edges labelled $1,2,...,N$. 
On each face of this domain of the square lattice, we draw at random, say with respective
probabilities $1-t_i,t_i$ in the $i$-th column (at the vertical of the point labelled $i$) one of the two 
following configurations
\eqn\threefaces{\vcenter{\hbox{\epsfxsize=1.2cm\epsfbox{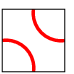}}} 
\qquad \qquad  \qquad\vcenter{\hbox{\epsfxsize=1.2cm\epsfbox{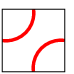}}}}
The strip is moreover supplemented with the following pattern of fixed configurations of
loops on the (left and right) boundaries:
\eqn\boundabro{\epsfxsize=6.cm\vcenter{\hbox{\epsfbox{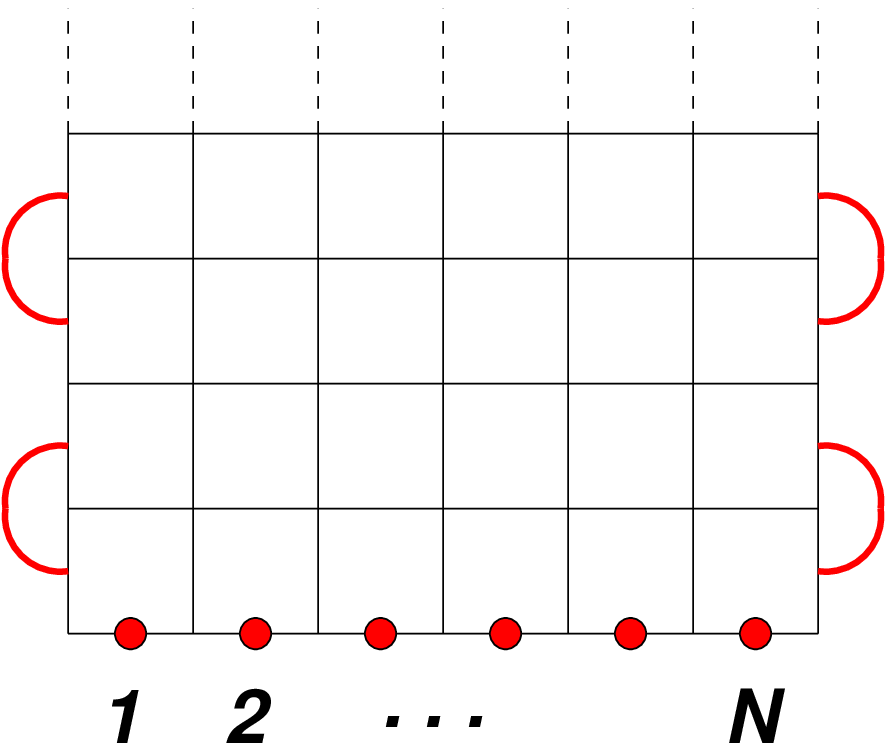}}}}
\fig{A sample configuration of the Dense Loop model on a strip
of width $N=6$ (left). We have indicated the corresponding open non-crossing link pattern of connection
of the points $1,2,3,4,5,6$ (right).}{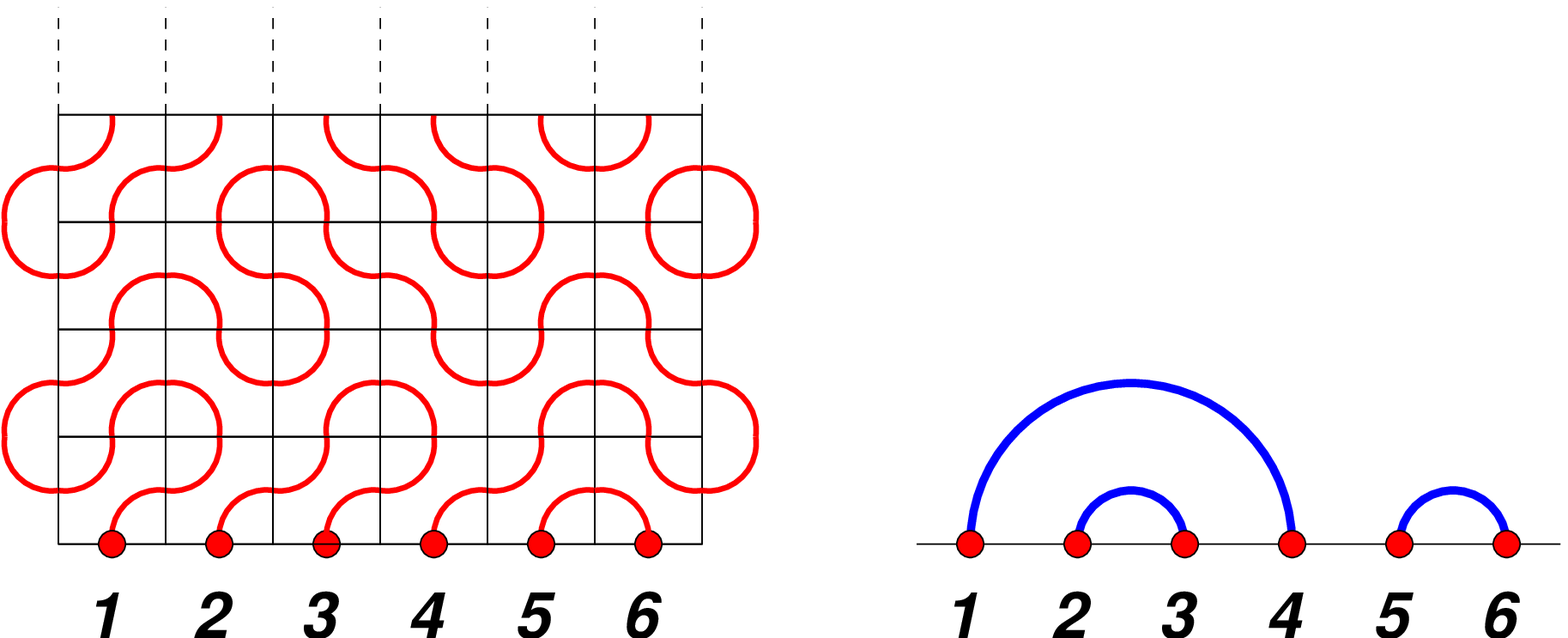}{12.cm}
\figlabel\stripop
With probability 1, a configuration will lead to a pairing of the points $1,2,...,N$ according
to their connection via the paths
(except for one point if
$N$ is odd which is connected to the infinity along the strip). Such a pattern
of connections is called  a {\it link pattern}, and an individual pairing is called an arch. 
The set of link patterns 
on $N$ points is denoted by 
$LP_{N}$, and has cardinality $(2n)!/(n!(n+1)!)$ for $N=2n$ or $N=2n-1$. 
Each link pattern of odd size $N=2n-1$ may indeed be viewed as a link pattern of size $2n$
but with the point $2n$ sent to infinity on the strip: this provides a natural bijection between
$LP_{2n-1}$ and $LP_{2n}$.
A link pattern $\pi\in LP_N$ may also be viewed as a permutation
$\pi \in S_N$ with only cycles of length 2 (except one cycle of length 1 for $N$ odd), 
and we shall use the notation $\pi(i)=j$
to express that points $i$ and $j$ are connected by an arch. For a pair $(i,j)$ such that
$j=\pi(i)$ and $i<j$, we will call $i$ the opening and $j$ the closing of the arch connecting $i$ and $j$.
An example of loop configuration
together with its link pattern are depicted in Fig.~\stripop. We use a standard pictorial representation
for link patterns in the form of configurations of non-intersecting arches connecting regularly spaced points on
a line, within the upper-half plane it defines. For odd $N$, the unmatched point
may be represented as connected to infinity in the upper-half plane via a vertical half-line. 
We moreover attach a weight $\tau=-(q+q^{-1})=1$ to each loop (hence
the denomination O($\tau=1$) model, $q=-e^{i\pi/3}$).
We may then compute the probability ${\rm Prob}(\pi)$
for a given randomly generated configuration of the loop
model on the strip to be connecting the boundary points according to a given link pattern $\pi$.

In Ref.~\DFOP, the model was solved by means of a transfer matrix technique, using solutions
of the boundary Yang--Baxter equation \SKLY\ \JAPOP\ that parametrize the inhomogeneous
probabilities $t_i$ via integrable Boltzmann weights, coded by a standard trigonometric 
$R$-matrix. Using the integrability of the system, and following the philosophy of \DFZJ, 
the suitably renormalized vector of probabilities $\Psi\equiv \{ \Psi_\pi \}_{\pi\in LP_N}$
was shown to satisfy the quantum Knizhnik--Zamolodchikov equation with reflecting 
boundaries for $q=-e^{i\pi/3}$, in the link pattern basis. 

In the following, we will consider the more general case of
generic $q$, $\tau$, which does not have stricto sensu an interpretation in terms of
lattice loop model \DFBQKZ.

\subsec{$R$ matrix}
The $R$-matrix of the model is an operator acting on link patterns
of $LP_N$:
\eqn\checR{\eqalign{\Rc_{i,i+1}(z,w)&={q z-q^{-1}w\over q w-q^{-1}z} \, \epsfxsize=1.2cm\vcenter{\hbox{\epsfbox{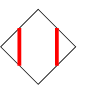} }}
+{z-w \over q w-q^{-1}z}\, \vcenter{\hbox{\epsfxsize=1.2cm\epsfbox{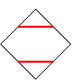} }}\cr
&={q z-q^{-1}w\over q w-q^{-1}z} I  +{z-w \over q w-q^{-1}z}\, e_i \cr}}
where $z$ and $w$ are complex numbers attached to the points $i$ and $i+1$ and
where $e_i$, $i=1,...,N-1$ are the generators of the Temperley-Lieb algebra $TL_N(\tau)$, subject
to the relations
\eqn\tla{ e_i^2=\tau e_i,\qquad [e_i,e_j]=0\ {\rm if}\ |i-j|>1,\qquad e_ie_{i\pm 1}e_i=e_i}
with the parametrization
\eqn\paratau{ \tau=-q-q^{-1} }
\fig{Action of the Temperley-Lieb generators $e_i$ on link patterns. }{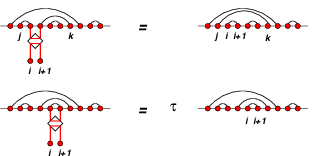}{11.cm}
\figlabel\actl
In \checR, we have depicted the Temperley-Lieb generators as tilted squares with
edge centers connected by pairs.
The corresponding action on link patterns should be understood as follows (see Fig.~\actl): 
assume the points $i,i+1$ 
are connected to say the points $j,k$ in a link pattern $\pi$. Then, unless $j=i+1$ and $k=i$,
the link pattern $\pi'= e_i \pi$ is identical to $\pi$
except that $i$ is now connected to $i+1$, and $j$ to $k$. If $j=i+1$ and $k=i$, the points
$i,i+1$ are connected to each other in $\pi$, and $e_i \pi =\tau \pi$. The latter is a direct consequence of the projector condition $e_i^2=\tau e_i$, as any link pattern with $i$ 
connected to $i+1$ lies in the image of $e_i$.

The above $R$-matrix satisfies the Yang--Baxter equation and the unitarity relation
\eqn\ybeunit{\eqalign{ \Rc_{i,i+1}(z,w)\Rc_{i+1,i+2}(z,x)\Rc_{i,i+1}(w,x)&=
\Rc_{i+1,i+2}(w,x)\Rc_{i,i+1}(z,x)\Rc_{i+1,i+2}(z,w)\cr
\Rc_{i,i+1}(z,w) \Rc_{i,i+1}(w,z)&=I \cr}}
as consequencs of the Temperley-Lieb algebra relations \tla.

\subsec{Link patterns, dyck paths, and containment order}
\fig{Dyck path (b) associated to a link pattern (a). The former is obtained
as the discrete path on the non-negative integer line: $0,1,2,1,2,3,2,1,0,1,0$.
%taking an $i$-th step up (resp.\ down) whenever an arch opens (resp. closes) at $i$.
We have represented in $(c)$ the box decomposition of the Dyck path.}{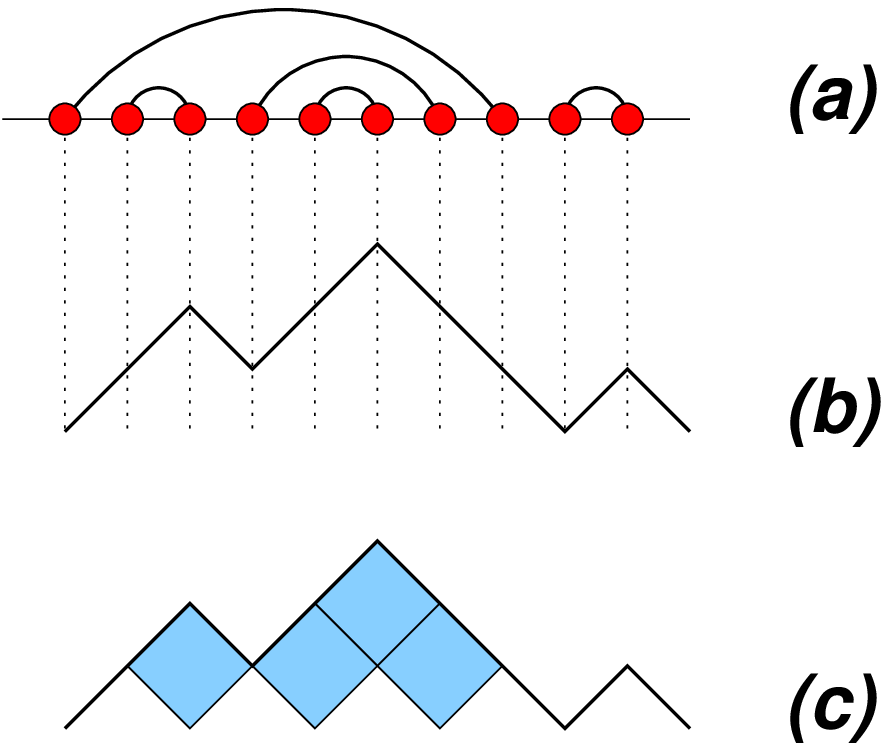}{6.cm}
\figlabel\lipadick

Before turning to the $q$KZ equation, we wish to emphasize a number of useful properties
satisfied by the link patterns, and the action of the Temperley-Lieb generators. An alternative
pictorial representation of link patterns is via Dyck paths, namely paths from and to
the origin on the non-negative integer line, with steps of $\pm 1$ only.
The bijection between link patterns and Dyck paths is illustrated on Fig.~\lipadick\ for $N=10$.
For even $N=2n$, we construct the Dyck path by visiting the points of the link pattern
from $1$ to $N$, starting from the origin of the non-negative integer half-line,
and with the following rule: if an arch opens (resp.\ closes) at $i$, then
the path goes up (resp.\ down) one step between time $i-1$ and $i$.
The path is guaranteed to come back to the origin at time $2n$
as there are as many openings as closings of arches, and moreover stays in the non-negative
half-line as all arches must first open before closing. In the case of odd $N=2n-1$,
one arch exactly has an opening and no closing point (it is connected to infinity), hence the 
path ends up at the point $1$ on the integer half-line. It can be completed into a path of length
$2n$ by a final step to the origin, thus expressing on Dyck paths the abovementioned
bijection between $LP_{2n-1}$ and $LP_{2n}$. The Dyck path is represented 
in the plane as the (broken-line) graph of the function $(t,h(t))$, $t=0,1,...,N$.

Dyck paths allow to endow the set of link patterns with a natural ``containment'' order,
namely $\pi < \pi'$ iff the Dyck path of $\pi$ contains strictly that of $\pi'$. This notion is
made explicit by introducing the ``box decomposition'' of any given Dyck path
(see Fig.~\lipadick\ (c)), namely the decomposition of the region of the plane delimited
by the path and a broken line $(0,0)\to (1,1)\to (2,0)\to \cdots \to (N-1,1)\to (N,0)$ if $N=2n$,
without the last step for $N=2n-1$, by use of squares of edge $\sqrt{2}$ tilted
by $45^\circ$.
Then a Dyck path $\delta$ contains strictly another $\delta'$ iff $\delta$ is obtained from 
$\delta'$ by addition of at least one box.
A box addition at position $m$ consists simply in replacing a portion of path 
$(m-1,h)\to (m,h-1)\to (m+1,h)$ that visits successively the points $h,h-1,h$ of the
integer half-line at times $m-1,m,m+1$, 
by the portion $(m-1,h)\to (m,h+1)\to (m+1,h)$, thus adding a
box with center at coordinates $(m,h)$. This may also be described as
transforming a local minimum into a local maximum at position $m$ on the path.
The ``smallest'' link pattern (whose Dyck path contains all others) is the pattern $\pi_0$
with links $\pi_0(i)=2n+1-i$, $i=1,2,...,n$ for $N=2n$, and $i=2,...,n$, for $N=2n-1$, while
$\pi_0(1)=1$. It corresponds to the farthest excursion, reaching point $n$ on the integer
half-line.  The ``largest'' link pattern (whose Dyck path is contained in all others)
$\pi_{\rm max}$ has $\pi_{\rm max}(i)=i+1$ for $i=1,3,..,2n-1$ when $N=2n$ and
$i=1,3,...,2n-3$ when $N=2n-1$, while $\pi_{\rm max}(2n-1)=2n-1$. It corresponds to
the shortest range excursion, alternating betwen the origin and point $1$ on the
integer half-line. So we have $\pi_0<\pi<\pi_{\rm max}$
for all $\pi\in LP_N$ such that $\pi\neq \pi_0,\pi_{\rm max}$. Finally, we shall denote
by $\beta(\pi)$ the total number of boxes in the box decomposition of the Dyck path
associated to $\pi$. We have for instance 
$\beta(\pi_0)=n(n-1)/2$ and $\beta(\pi_{\rm max})=0$.

\fig{Box addition at position $i$ on Dyck paths corresponding to the action of $e_i$. We have depicted three generic situations for the addition: (i) at a local minimum (ii) at a local maximum (iii) at a slope. Both (ii) and
(iii) lead to a Dyck path contained by the original one, while (i) produces a Dyck path containing it,
with exactly one additional box.}{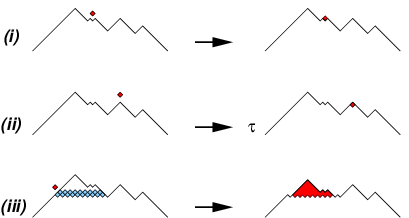}{13.cm}
\figlabel\troisitu
The action of $e_i$ on link patterns
may be easily translated into the language of boxes on Dyck paths. The action of $e_i$
may indeed be viewed as a box addition at position $i$ on the corresponding Dyck paths.
Then 3 situations may occur (Fig.~\troisitu):
\item{(i)} The path has a minimum at point $i$: the box addition transforms it into a maximum.
\item{(ii)} The path has a maximum at point $i$: the box-added path is unchanged, but picks up a factor of $\tau$.
\item{(iii)} The path has a slope at $i$, namely a succession of two up or two down steps: the box addition actually destroys the two rows of boxes at its height and immediately below 
until the other side of the path is reached. The net result may be interpreted as an avalanche,
in which the mountain shape between the point of impact and the other side falls down by two units.
\par
This allows to see that among all possible actions of $e_i$ on a link pattern $\pi$, only one
leads to a ``larger'' Dyck path (containing $\pi$): $e_i\pi=\pi'<\pi$, namely in the situation $(i)$, while all other 
situations lead to $\pi<\pi'=e_i \pi$. This observation will be used below.

The interpretation of the action of $e_i$ on Dyck paths was used in \PRdG\ to 
rephrase the homogeneous loop model as the stochastic model of a growing interface.

\newsec{The $q$KZ equation for reflecting boundary condition}
\subsec{The equation}
The reflecting boundary $q$KZ equation consists of 
the following system of equations for a vector $\Psi$ which depends polynomially on the
variables $z_1,\ldots,z_N$ (and $q,q^{-1}$):
\eqna\qkz
$$\eqalignno{
\Rc_i(z_{i+1},z_i)\Psi_N(z_1,\ldots,z_i,z_{i+1},\ldots,z_N)
&=
\Psi_N(z_1,\ldots,z_{i+1},z_i,\ldots,z_N)
&\qkz{a}\cr
c_N(z_N)\Psi_N(z_1,\ldots,z_N)
&=
\Psi_N(z_1,\ldots,z_{L-1},s/z_N)
&\qkz{b}\cr
c_1(z_1)\Psi_N(z_1,\ldots,z_N)
&=
\Psi_N(1/z_1,z_2,\ldots,z_N)
&\qkz{c}\cr
}$$
Here $c_1$ and $c_N$ are scalar functions to be specified later, and $s=q^{2(k+2)}$ is a parameter which determines the ``level'' $k$ of the equation: here we consider the so-called level $1$ case, 
namely with $s=q^6$.

One can think of Eqs.~\qkz{} as an analogue of the quantum 
Knizhnik--Zamolodchikov equation ($q$KZ)
in the form introduced by Smirnov \Smi\ (see also \FR),
in which one replaces the periodic boundary conditions, implicit in the usual $q$KZ, with reflecting 
boundaries \JAPOP.
More precisely, Eq.~\qkz{a} is the exchange relation corresponding to the bulk, independent
of boundary conditions, whereas Eqs.~\qkz{b,c} implement the reflections at the two boundaries.

In \DFBQKZ, it was remarked that solving these equations for even size $N=2n$ automatically provides 
a solution for odd size $N-1$ by taking the last parameter $z_N$ to zero (or equivalently to infinity).
We therefore discuss in detail the case of even size now, postponing to Sect.~6
the discussion of the odd case.

\subsec{Minimal polynomial solution}
In \DFOP, it was claimed that the system of equations \qkz{} possesses a polynomial solution of minimal degree
$3n(n-1)$ which is unique up to multiplication by a scalar. To actually solve the equations \qkz{},
one first remarks that the $N-1$ equations \qkz{a} from a triangular system with respect
to the containment order of Dyck paths introduced in Sect.~2.3. Indeed, when written in
components, this equation reads:
\eqn\compopsi{{q^{-1}z_{i+1}-q z_i\over z_{i+1}-z_i} (\tau_i -1)\Psi_\pi(z_1,\ldots,z_{2n}) 
=\sum_{\scriptstyle\pi': \pi'\neq \pi\atop\scriptstyle e_i \pi'=\pi} 
\Psi_{\pi'}(z_1,\ldots,z_{2n}) }
where $\tau_i$ acts on functions of the $z$'s by interchanging $z_i$ and $z_{i+1}$.
Now consider the sum on the r.h.s.: it extends over the proper inverse images of $\pi$
under $e_i$. Picking $\pi$ in the image of $e_i$ (i.e. with an arch connecting points
$i$ and $i+1$, as explained above), its inverse images $\pi'$ under $e_i$ all have dyck paths
containing that of $\pi$ (i.e.\ $\pi'<\pi$) except one, say $\pi^*$, corresponding to the Dyck path
of $\pi$ with the box at position $i$ removed, hence with $\pi<\pi^*$. Hence Eq.~\compopsi\
allows to express $\Psi_{\pi^*}$ in terms of only $\Psi_\alpha$, with $\alpha<\pi^*$. 
The solution is therefore uniquely fixed by specifying the component
corresponding to the smallest link pattern $\pi_0$ defined above,
whose Dyck path that contains all others. The latter is entirely fixed by the degree condition and factorization
properties deduced from the $q$KZ system; the result is:
\eqn\base{
\Psi_{\pi_0}=\prod_{1\le i<j\le n} (q\,z_i-q^{-1}z_j)(q^2-z_iz_j)
\prod_{n+1\le i<j\le 2n} (q\,z_i-q^{-1}z_j)(q^4-z_iz_j)
}
This fixes the functions $c_1$ and $c_N$ in Eqs.~\qkz{b,c} to be $c_1(x)=1/x^{2n-2}$ and 
$c_N(x)=(q^3/x)^{2n-2}$.

It is also a simple exercise to prove that the solution enjoys a reflection invariance 
property, inherited from the inversion (unitarity) relation satisfied by the $R$-matrix \ybeunit,
and clearly satisfied by the fundamental component $\Psi_{\pi_0}$:
\eqn\inversion{ \Psi_\pi(z_1,\ldots,z_{2n})=\prod_{i=1}^{2n}\left({z_i\over q^3}\right)^{n-1}
\Psi_{\rho(\pi)}\left( {q^3\over z_{2n}},\ldots,{q^3\over z_1}\right) }
where $\rho(\pi)$ is the mirror image of $\pi$ w.r.t.\ a vertical axis. 

At the special value $q=-\exp(i\pi/3)$, which is a ``degenerate'' case of the $q$KZ system where the shift $s$ becomes
unity, the vector $\Psi$ can be interpreted as the ground state eigenvector of the loop model described in Sect.~2.
For generic $q$, no such direct physical interpretation of $\Psi$ exists; however, it still retains remarkable
combinatorial properties that we shall describe now.

\subsec{Conjectures}
In Ref.~\DFVS, a number of conjectures were made on the homogeneous, generic $q$ limit
of the components of $\Psi$. The dependence on $q$ was always expressed through
the quantity $\tau=-q-q^{-1}$, and the entries of $\Psi$ were observed to be 
polynomials of $\tau$ with non-negative integer coefficients. The main conjectures concerned an identification of the sum
of components of $\Psi$ with generating polynomials
for weighted rhombus tilings of finite domains of the triangular lattice, involving the catalytic 
variable $\tau$. These were interpreted in odd and even sizes as $\tau$-enumerations of respectively
Cyclically Symmetric Transpose Complementary Plane Partitions (CSTCPP), or equivalently
of rhombus tilings of a hexagon with cyclic and transpose-complementary symmetries (odd size),
and in the even case as modified CSTCPP (CSTCPP$^\triangle$)
corresponding to rhombus tilings of a 
hexagon with a central triangular hole and with cyclic and transpose-complementary symmetries.
These conjectures were extensions of earlier conjectures concerning only the particular value $\tau=1$
($q$ cubic root of unity), due to Razumov and Stroganov \OPRS, and proved in \DFOP,
involving respectively the total number of CSTCPP in odd size, and that of Vertically Symmetric Alternating Sign Matrices (VSASM) in even size.

Apart from the main sum rule conjectures, a number of other conjectures were made in \DFVS\
on the entries of $\Psi$ as polynomials of $\tau$, concerning degree and valuation, 
and also conjecturally
relating the small $\tau$ behavior to the numbers of Totally Symmetric Self-Complementary Plane 
Partitions (TSSCPP) with fixed features, namely, once expressed as Non-Intersecting Lattice Paths (NILP), with fixed termination points of the paths.

The purpose of the present paper is to prove these various conjectures, by means of multiple integral
expressions for the homogeneous solution to the $q$KZ equation $\Psi(\tau)$.

\newsec{Integral expressions for solutions of level $1$ $q$KZ}
\subsec{The intermediate basis}
The method introduced in \ZJDF\ in order to obtain integral representations of $\Psi$ was to exhibit a different basis
than that of link patterns in which the integral expressions for the components are relatively
simple. We shall use the same basis here. Note that this section is ``boundary conditions-independent'' and its results
are equally valid for say periodic boundary conditions.

The elements of the basis we consider are indexed by strictly
increasing integer sequences of the form ${\bf a} \equiv \{a_1,a_2,\ldots,a_n\}$, where $1\leq a_i\leq 2i-1$.
We denote by $O_n$ the set of such sequences.
The sequences in $O_n$ are in one-to-one correspondence with the link patterns in $LP_{2n}$ 
in two different, inequivalent ways. One may indeed associate to each $\pi\in LP_{2n}$ the sequence
$a_i(\pi)$, $i=1,2,\ldots,n$, of the positions (counted from left to right, and taking values 
in $\{1,2,\ldots,2n-1\}$) of the openings of arches in $\pi$. Similarly, we may associate to
$\pi$ the sequence $b_i(\pi)=a_i(\rho(\pi))$, $i=1,2,\ldots,n$ recording the closing positions
of the arches of $\pi$, counted from right to left, or equivalently the opening positions of the
arches in the reflection $\rho(\pi)$.

In \ZJDF, we have constructed the change of basis from the link pattern basis to the ``arch opening''
basis, namely expressed the solutions $\Psi_\pi$, $\pi\in LP_{2n}$ in terms of multiple residue integrals 
$\Psi_{a_1,\ldots,a_n}$, with $\{a_1,a_2,\ldots,a_n\}\in O_n$. More precisely, Ref.~\ZJDF\ expresses any 
integral $\Psi_{\bf a}\equiv \Psi_{a_1,\ldots,a_n}$ for weakly increasing sequences of $a$'s as well, in terms
of the solution $\Psi_\pi$ in the link pattern basis, via the linear transformation:
\eqn\matrirela{ \Psi_{\bf a}(z_1,\ldots, z_{2n})=\sum_{\pi\in LP_{2n}}  C_{{\bf a},\pi}(\tau) \Psi_\pi(z_1,\ldots, z_{2n})}
with polynomial coefficients $ C_{{\bf a},\pi}(\tau)$ expressed as follows. 
Let $U_k\equiv (q^{k+1}-q^{-k-1})/(q-q^{-1})$. For $k\ge 0$,
$U_k$ is the Chebyshev polynomial of the second kind associated to $-\tau$, defined
recursively by $U_{k+1}=-\tau U_k -U_{k-1}$ with $U_0=1$ and $U_{-1}=0$. A given link pattern $\pi \in LP_{2n}$
may alternatively be thought of as a permutation of $\{1,2,\ldots,2n\}$ with only cycles of length $2$. The arches 
forming $\pi$ may therefore be described by the pairs $(i,\pi(i))$, $1\leq i\leq 2n$, such that $i<\pi(i)$, in which case
$i$ are the positions of the openings and $\pi(i)$ of the closings of the arches in $\pi$.
Then we have the following formula:
\eqn\formulA{ C_{{\bf a},\pi}(\tau)=\prod_{\scriptstyle i=1\atop\scriptstyle i<\pi(i)}^{2n-1} U_{\mu({\bf a},i)} }
where 
\eqn\defmu{ \mu({\bf a},i) = {\rm card}\left\{ j \vert i \leq a_j <\pi(i) \right\} +{i-\pi(i)-1\over 2} }
In \ZJDF, it was mentioned that the entries \formulA\ may easily be computed by iteratively removing the 
``little arches'' of $\pi$ (with $\pi(i)=i+1$), and replacing them with a factor $U_{m-1}$, 
where $m$ is the total number of $a$'s lying
under that arch (namely such that $a_j=i$). 
The new link pattern thus obtained has one less arch, and its $a$'s are relabeled
accordingly, while $m-1$ extra $a$'s are placed in position $i-1$. 
The algorithm is iterated until the link pattern becomes empty.

If we moreover restrict the set of $a$'s to $O_n$, we get a true change of basis from link patterns to arch openings,
in which $C(\tau)$ is a square invertible $c_n \times c_n$ matrix with polynomial coefficients. The matrix $C(\tau)$ indeed 
enjoys the following properties. Let us first use the bijection between $LP_{2n}$ and $O_n$ to write 
$C_{{\bf a},\pi}(\tau)\equiv C_{\alpha,\pi}(\tau)$, where $\alpha\in LP_{2n}$ is uniquely determined by its arch opening
positions $a_1,a_2,\ldots a_n$, counted from left to right. Let us moreover order the link patterns, say by increasing lexicographic order
on the sets of positions of their arch openings. Then we have the property
\item{({\cal P})} $C(\tau)$ is a lower triangular matrix, with entries $1$ on the diagonal, and polynomials of $\tau$ with integer
coefficients elsewhere, and the same holds for $C^{-1}(\tau)$.

\par\noindent Property ({\cal P}) is easily derived as follows. 
First, it is clear that the diagonal terms $C_{{\bf a}(\pi),\pi}=1$, as $\mu({\bf a}(\pi),i)=0$
for all arch openings $i$ of $\pi$: indeed, the $a$'a being the arch openings of $\pi$, the set $\{  a_j \vert i \leq a_j <\pi(i) \}$ 
has one $a$ per arch enclosed by the arch $i\to \pi(i)$, henceforth a total of $(\pi(i)-i+1)/2$. Consequently, all the indices
of the Chebyshev polynomials contributing to \formulA\ vanish, and as $U_0=1$, the result follows. The triangularity is best
understood by following the abovementioned algorithm for constructing $C$. Indeed, at any step in the algorithm, the only cause
for the matrix element to vanish is that one has no $a$'s under the little arch considered, as in this case one would get a factor $U_{-1}=0$.
The matrix element $C_{\alpha,\pi}$ can only be non-zero if the arch openings of $\alpha$ occupy positions lexicographically larger that those of $\pi$. Indeed,
by contradiction, assume the structure of arch openings in $\alpha$ and $\pi$ are the same up to say a position $i$ where an arch opens
 in $\alpha$ while an arch closes in $\pi$. This means that the total number of arch openings for positions $j>i$ is strictly larger in $\pi$ 
 than in $\alpha$. Consequently,  applying the above algorithm to the arches of $\pi$ opening at positions $> i$,
 we see that at least one little arch in the process will have no arch opening of $\alpha$ below it, 
 thus receiving a weight $U_{-1}=0$, and therefore the corresponding matrix element $C_{\alpha,\pi}$ vanishes.

Finally, one can rewrite the $q$KZ equation itself using the linear combinations defined by Eq.~\matrirela. Note here that
we are forced to use not only the components corresponding to our basis $O_N$ of increasing sequences, but also those
corresponding to any non-decreasing sequence. 
In principle all of them can be reexpressed as linear combinations of increasing sequences only, but it is
preferrable to avoid having to write these linear dependence relations explicitly.

All that is needed is the action of the $e_i$ on the $\Psi_{\bf a}$. We have the following

\noindent{\bf Theorem 1:}
{\sl For any non-decreasing sequence $a_1,\ldots,a_n$ such that the number
$i$ occurs exactly $k$ times, $k\ge0$, we have the formula:
\eqnn\eibasea
$$\eqalignno{
(e_i \Psi)_{a_1,\ldots,\underbrace{\scriptstyle i,\ldots,i}_{k},\ldots,a_n}=&
U_{k-1}U_{k-4}
\Psi_{a_1,\ldots,a_n}
-U_{k-1}U_{k-3}
\big(\Psi_{a_1,\ldots,i-1,\underbrace{\scriptstyle i,\ldots,i}_{k-1},\ldots,a_n}\cr
&+\Psi_{a_1,\ldots,\underbrace{\scriptstyle i,\ldots,i}_{k-1},i+1,\ldots,a_n}\big)
+U_{k-1}U_{k-2}
\Psi_{a_1,\ldots,i-1,\underbrace{\scriptstyle i,\ldots,i}_{k-2},i+1,\ldots,a_n}&\eibasea
}$$
(where for $k=0$ the r.h.s.\ is zero).}

\noindent{}Proof: expand the l.h.s.\ in the basis of link patterns by using Eq.~\matrirela. We find:
\eqn\eibaseaexp{
(e_i \Psi)_{\bf a}=\sum_{\pi: \pi(i)\ne i+1} C_{\bf a,e_i(\pi)} \Psi_\pi
+\tau\sum_{\pi: \pi(i)= i+1} C_{\bf a,\pi} \Psi_\pi
}
where we have distinguished among the entries of $e_i$ its diagonal entries, equal to $\tau$, and its non-diagonal entries,
equal to $1$.

The $\Psi_\pi$ must be regarded here as independent objects, so that 
we must now check Eq.~\eibasea\ for each link pattern $\pi$.
This will be performed by a case by case analysis of the situation around the sites $(i,i+1)$. Each time
only the coefficients involving the arches starting or ending at $i,i+1$ differ from term to term in the equation,
so that we can ignore the remaining factors.
The proof will be explained pictorially using the same conventions as in appendix
1 of \ZJDF, that is by drawing the coefficient $C_{\bf a,\pi}$ as the usual (local)
depiction of the link pattern $\pi$ decorated by placing between sites $i$ and $i+1$
(inside a circle) the total number $k$ of $a$'s such that $a_j=i$.

There are 4 cases:

\item{(i)} If $i$ is an opening and $i+1$ a closing of $\pi$, then $\pi$ has a little arch $(i,i+1)$: $\pi(i)=i+1$.
In this case the equality reduces pictorially to
$$\eqalign{
\tau\vcenter{\hbox{\epsfxsize=3cm\epsfbox{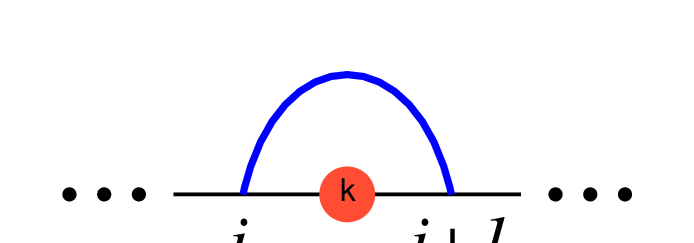}}}
&=U_{k-1}U_{k-4} \vcenter{\hbox{\epsfxsize=3cm\epsfbox{caseA1.eps}}}
-U_{k-1}U_{k-3} \vcenter{\hbox{\epsfxsize=3cm\epsfbox{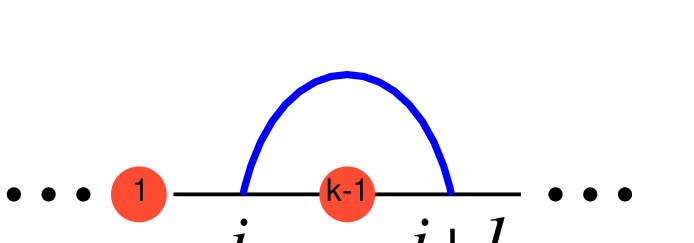}}}\cr
&-U_{k-1}U_{k-3} \vcenter{\hbox{\epsfxsize=3cm\epsfbox{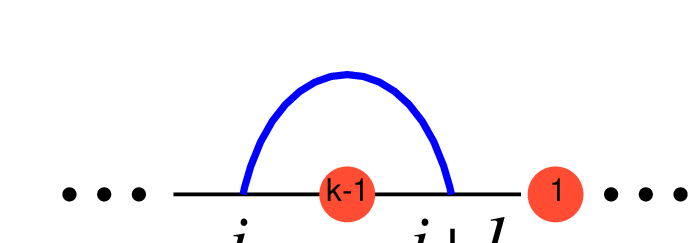}}}
+U_{k-1}U_{k-2} \vcenter{\hbox{\epsfxsize=3cm\epsfbox{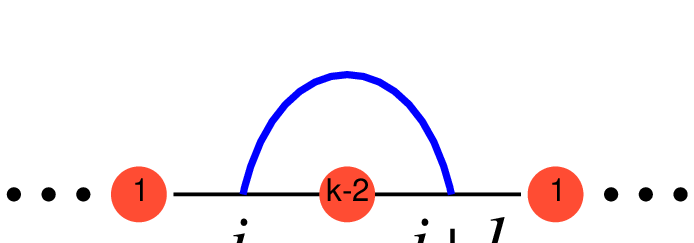}}}
}$$
or explicitly
\eqn\eibaseaA{
\tau U_{k-1}=
U_{k-1}U_{k-4}\times U_{k-1}-U_{k-1}U_{k-3}\times 2 U_{k-2}+U_{k-1}U_{k-2}\times U_{k-3}
}
which is easily checked by noting that $U_{k-1}U_{k-4}-\tau=U_{k-2}U_{k-3}$.

In all other cases there is no little arch $(i,i+1)$.

\item{(ii)} If both $i$ and $i+1$ are openings, call $p$ the total ``weight'' under the arch leaving $i+1$,
that is $p={\rm card}\{\ell|i+1\le a_\ell<\pi(i+1)\}-{i+1-\pi(i+1)-1\over 2}$,
and $q$ the remaining weight under the bigger arch starting from $i$, 
excluding what is under the smaller arch and the weight $k$ under the segment $[i,i+1)$, 
in order to make the pictorial
description simpler: $q={\rm card}\{\ell|\pi(i+1)\le a_\ell<\pi(i)\}-{\pi(i)-\pi(i+1)-1\over 2}$.
Then the identity to prove is:
$$\eqalign{
\vcenter{\hbox{\epsfxsize=3cm\epsfbox{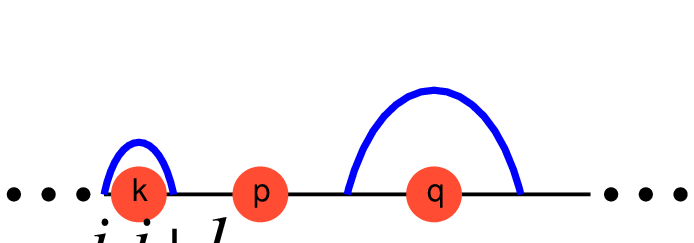}}}
&=U_{k-1}U_{k-4} \vcenter{\hbox{\epsfxsize=3cm\epsfbox{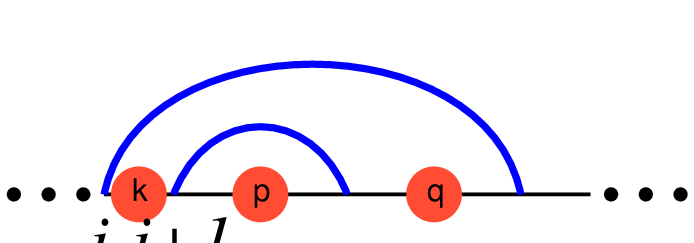}}}
-U_{k-1}U_{k-3} \vcenter{\hbox{\epsfxsize=3cm\epsfbox{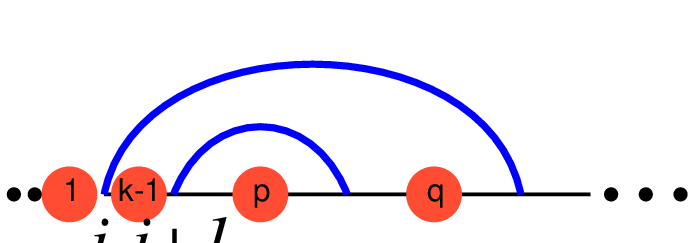}}}\cr
&-U_{k-1}U_{k-3} \vcenter{\hbox{\epsfxsize=3cm\epsfbox{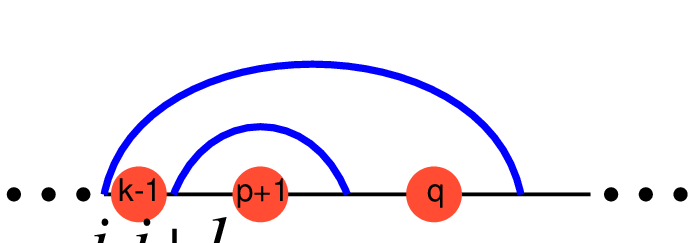}}}
+U_{k-1}U_{k-2} \vcenter{\hbox{\epsfxsize=3cm\epsfbox{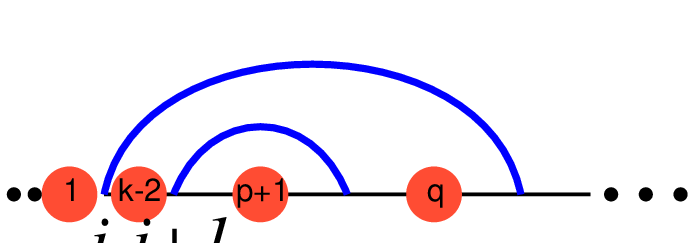}}}
}$$
\eqnn\eibaseB
$$\eqalignno{
U_{k-1}U_{q-1}=&
U_{k-1}U_{k-4}\times U_{p-1}U_{k+p+q-2}
-U_{k-1}U_{k-3}\times U_{p-1}U_{k+p+q-3}\cr
&-U_{k-1}U_{k-3}\times U_{p}U_{k+p+q-2}
+U_{k-1}U_{k-2}\times U_{p}U_{k+p+q-3}
&\eibaseB\cr}
$$
which is again a routine check.

\item{(ii')} The case where $i$ and $i+1$ are both closings is treated analogously.

\item{(iii)} Finally, if $i$ is a closing and $i+1$ an opening,
call $p$ the weight under the arch $(\pi(i),i)$ defined as before, 
and $q$ the weight under the arch $(i+1,\pi(i+1))$. Similary the proof of the identity
$$\eqalign{
\vcenter{\hbox{\epsfxsize=3cm\epsfbox{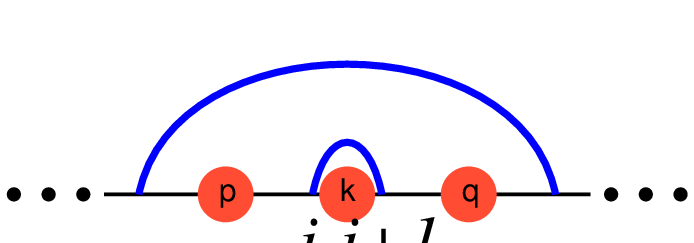}}}
&=U_{k-1}U_{k-4} \vcenter{\hbox{\epsfxsize=3cm\epsfbox{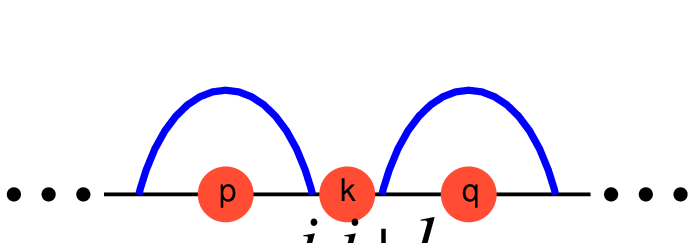}}}
-U_{k-1}U_{k-3} \vcenter{\hbox{\epsfxsize=3cm\epsfbox{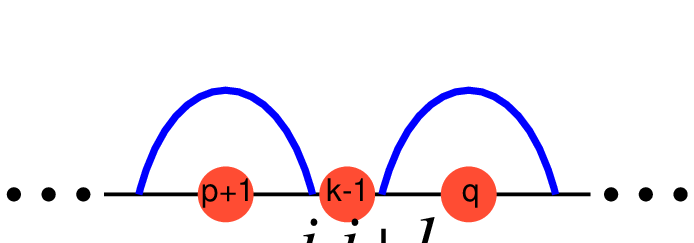}}}\cr
&-U_{k-1}U_{k-3} \vcenter{\hbox{\epsfxsize=3cm\epsfbox{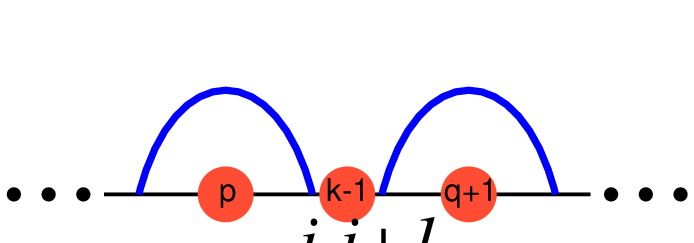}}}
+U_{k-1}U_{k-2} \vcenter{\hbox{\epsfxsize=3cm\epsfbox{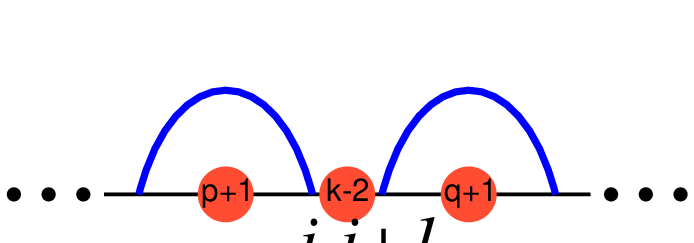}}}
}$$
\eqnn\eibaseC
$$\eqalignno{
U_{k-1}U_{k+p+q-2}=&
U_{k-1}U_{k-4}\times U_{p-1}U_{q-1}
-U_{k-1}U_{k-3}\times U_p U_{q-1}  \cr
&-U_{k-1}U_{k-3}\times U_{p-1} U_q
+U_{k-1}U_{k-2}\times U_p U_q
&\eibaseC\cr}
$$
is left to the reader. This completes the proof of the theorem 1.

\subsec{Integral solution of $q$KZ equation: general principle}
The idea to use integral representations for solutions of the $q$KZ equation is not new
and there is a vast literature on the subject (cf the references in Sect.~11.2 of \JMbook).
We consider here a very specific type of level 1 solutions, for which one expects a much simpler formula
than generically. In the present context, this idea was used in \ZJDF\ in the case of the $q$KZ equation
with the usual periodic boundary conditions. We now describe the procedure in a slightly more general 
(boundary conditions-independent) setting.

The idea is to define for any non-decreasing sequence $(a_1,\ldots,a_n)$ the following quantity:
\eqnn\generalpsi
$$\eqalignno{ \Psi_{a_1,\ldots,a_n}&(z_1,\ldots,z_N)=
\prod_{1\leq i<j\leq N} (q z_i-q^{-1}z_j)
\oint\cdots\oint \prod_{m=1}^n  
{d w_\ell\over 2\pi i}\cr
&
F(z_1,\ldots,z_N;w_1,\ldots,w_n)
{\prod_{1\leq \ell<m\leq n} (w_m-w_\ell)(q w_\ell-q^{-1}w_m)
\over
\prod_{\ell=1}^n
\prod_{i=1}^{a_\ell}(w_\ell-z_i)\prod_{i=a_\ell+1}^{N}
( q w_\ell -q^{-1}z_i)}
&\generalpsi
\cr}
$$
where $F$ is any rational function that is symmetric in all $z_i$ and symmetric in all $w_\ell$.
The contours of integration encircle the $z_i$ but not $q^{-2}z_i$, nor any poles of $F$.

We wish to prove that \generalpsi\ solves the exchange relation \compopsi\ of the $q$KZ equation,
in the operatorial form  $ t_i \Psi= (e_i-\tau)\Psi $ where the action of $e_i$ is given by Eq.~\eibasea,
and we have introduced he divided difference operator $t_i=(q z_i -q^{-1}z_{i+1}) \partial_i $. 
As before, we have to consider the cases where exactly $k$ $a$'s
take the value $i$, say $a_m=a_{m+1}=\cdots=a_{m+k-1}=i$, while $a_{m-1}<i$ or $m=1$ and $a_{m+k}>i$ or $m+k=N$. 
Starting from the expression \eibasea, and subtracting $\tau \Psi$ 
on both sides, we wish to express
$(e_i-\tau)\Psi$ as the action on $\Psi$ of the divided difference operator 
$t_i$. We are left with proving the following 

\noindent{\bf Theorem 2:} {\sl The function \generalpsi\ solves the exchange relation of the $q$KZ equation,
namely it satisfies:

\eqn\toprove{ \eqalign{
&(t_i\Psi)_{a_1,\ldots,a_{m-1},\underbrace{\scriptstyle i,\ldots,i}_{k},a_{m+k},\ldots,a_n}=
U_{k-2}U_{k-3}
\Psi_{a_1,\ldots,a_{m-1},\underbrace{\scriptstyle i,\ldots,i}_{k},a_{m+k},\ldots,a_n}\cr&
-U_{k-1}U_{k-3}
\big(\Psi_{a_1,\ldots,a_{m-1},i-1,\underbrace{\scriptstyle i,\ldots,i}_{k-1},a_{m+k},\ldots,a_n}
+\Psi_{a_1,\ldots,a_{m-1},\underbrace{\scriptstyle i,\ldots,i}_{k-1},i+1,a_{m+k},\ldots,a_n}\big)\cr
&+U_{k-1}U_{k-2}
\Psi_{a_1,\ldots,a_{m-1},i-1,\underbrace{\scriptstyle i,\ldots,i}_{k-2},i+1,a_{m+k},\ldots,a_n}\cr}}
}

\noindent{Proof}:
Two important remarks are in order.
Firstly, the operator $t_i$ acts only on the pieces of $\Psi$ that are non-symmetric in $(z_i,z_{i+1})$. When acting with $t_i$ on 
\generalpsi, we may restrict our attention to the non-symmetric part of the integrand. Secondly, we note that for any function
$S(u_1,\ldots,u_k)$ satisfying the following vanishing antisymmetrizer property that 
\eqn\antisym{ {\cal A}(S)\equiv \sum_{\sigma \in S_k} (-1)^\sigma S(u_{\sigma(1)},\ldots, u_{\sigma(k)}) =0}
then the multiple integral
\eqn\multipleS{I_k\equiv  \oint du_1\cdots du_k S(u_1,\ldots, u_k) \prod_{1\leq \ell<m \leq k} (u_m-u_\ell) =0}
Indeed, for any given permutation $\sigma\in S_k$, we may perform the change of variables $u_m=v_{\sigma(m)}$
for $m=1,2,\ldots,k$, resulting in 
$I_k=\oint d{\bf v} S({\bf v}^\sigma) \Delta({\bf v}^\sigma)=\oint d{\bf v}(-1)^\sigma S({\bf v}^\sigma)\Delta({\bf v})$,
where we have used the antisymmetry of the Vandermonde determinant, and finally summing over all permutations
yields \multipleS, thanks to \antisym. We also note that if $S$ satisfies \antisym, any symmetric function of the $u$'s 
multiplied by $S$ will also satisfy it.

As a consequence of the two above remarks, to prove the identity \toprove, it is sufficient to prove a weaker statement
on the part $P$ of the integrand of $\Psi$ that is non-symmetric in $(z_i,z_{i+1})$ and also non-symmetric
in $(w_{m},w_{m+1},\ldots,w_{m+k-1})$, after factoring out a Vandermonde determinant of the $w$'s. 
Rewriting $u_i=w_{m+i-1}$ for $i=1,2,\ldots,k$, $P$ reads simply
\eqn\defP{ P_k\equiv (q z_i-q^{-1}z_{i+1}){\prod_{1\leq \ell<m\leq k} (q u_\ell-q^{-1}u_m)
\over \prod_{\ell=1}^k (u_\ell-z_i)(q u_\ell-q^{-1}z_{i+1}) } }
Now introducing the quantity 
\eqn\introS{S(u_1,\ldots,u_k)\equiv   t_i P_k -
\Big(
U_{k-2}U_{k-3}
+U_{k-1}U_{k-3}(f_1+g_k)-U_{k-1}U_{k-2}
f_1g_k \Big) P_k
}
where we use the notations
\eqn\deffgs{ f_\ell= {u_\ell-z_i\over q u_\ell-q^{-1}z_i} \qquad g_\ell={q u_\ell-q^{-1}z_{i+1}\over u_\ell-z_{i+1}} }
for $\ell=1,2,\ldots,k$, we are simply left with the task of proving that \antisym\ is indeed satisfied, and \toprove\
will follow from the above considerations. 

For simplicity, we will work with the quantity 
$T=S\times {z_{i+1}-z_i\over q\,z_i-q^{-1}z_{i+1}}
\prod_{\ell=1}^k (u_\ell-z_{i})(q\, u_\ell-q^{-1}z_{i+1})$,
proportional to $S$ of \introS\ by a factor symmetric in the $u$'s, henceforth proving ${\cal A}(S)=0$ amounts to 
proving ${\cal A}(T)=0$. Explicitly, we have
\eqn\expliciST{ \eqalign{
{\cal A}&\Big(T(u_1,\ldots,u_k)\Big)= 
{\cal A}\Big(\Delta_q({\bf u})\Big) \left( (q z_{i+1}-q^{-1}z_{i})
\prod_{\ell=1}^k f_\ell g_\ell-(qz_{i}-q^{-1}z_{i+1})\right)\cr
&-(z_{i+1}-z_i){\cal A}\Big( \Delta_q({\bf u}) ( U_{k-2}U_{k-3}-U_{k-1}U_{k-3}(f_1+g_k)+U_{k-1}U_{k-2}f_1g_k)\Big) \cr}}
where we have written for short the $q$-Vandermonde as
$\Delta_q({\bf u})\equiv \prod_{1\leq \ell<m\leq k} (qu_\ell-q^{-1}u_m)$ and noticed that only this piece of $P$
is non-symmetric in the $u$'s.

We need the following lemma, expressing the antisymmetrization of the $q$-Vandermonde:
\eqn\antiqvander{ {\cal A} \Big( \Delta_q({\bf u})\Big) =(-1)^{k(k-1)/2} U_1U_2 \cdots U_{k-1} \ \Delta({\bf u}) }
Proof. We proceed by induction. For $k=2$, an explicit computation leads to
${\cal A}(q u_1-q^{-1}u_2)=(q+q^{-1})(u_1-u_2)=-U_1\Delta({\bf u})$.
Assume \antiqvander\ holds for $k\to k-1$.
We decompose any permutation $\sigma\in S_k$ according to the image of $1$, say $\sigma(1)=m$. Upon relabelling 
indices, the corresponding permutation $\sigma'$ is in $S_{k-1}$, and we may apply the recursion hypothesis 
to the summation over such $\sigma'$. We have
\eqn\antidecomp{\eqalign{ {\cal A} \Big( \Delta_q({\bf u})\Big)&= \sum_{m=1}^k 
\sum_{\scriptstyle\sigma\in S_k\atop\scriptstyle\sigma(1)=m} 
(-1)^\sigma \Delta_q({\bf u}^{\sigma})\cr
&=(-1)^{k(k-1)/2}\sum_{m=1}^k \prod_{\scriptstyle\ell=1\atop \scriptstyle\ell\neq m}^k 
{q u_m -q^{-1}u_\ell\over u_m-u_\ell} 
U_1U_2 \cdots U_{k-2} \Delta({\bf u})\cr}}
where we have applied the recursion hypothesis to $\Delta_q(u_1,\ldots,u_{m-1},u_{m+1},\ldots,u_k)$ and
reabsorbed the sign change by $(-1)^{k-1}$ into the ratio $\Delta({\bf u})/\prod_{\ell\neq m} (u_m-u_\ell)$.
To conclude, we still have to prove the following sublemma:
\eqn\sublemmauseful{ \varphi_k(u_1,\ldots,u_k)\equiv 
\sum_{m=1}^k \prod_{\scriptstyle\ell=1\atop \scriptstyle\ell\neq m}^k 
{q u_m -q^{-1}u_\ell\over u_m-u_\ell} =U_{k-1} }
for all distinct complex numbers $u_1,\ldots,u_k$.
To prove the latter, let us first note that it is a rational fraction, symmetric in the $u$'s. Viewed as a function of $u_1$,
it has possible poles at $u_2,u_3,\ldots,u_k$ and is bounded at infinity. By symmetry, it is sufficient to compute the
residue at $u_1\to u_2$, for which only the two first terms in the summation contribute, leading to:
\eqn\leadtovan{ {\rm Res}_{u_1\to u_2}\left({qu_1-q^{-1}u_2\over u_1-u_2} \prod_{\ell=3}^k {q u_1 -q^{-1}u_\ell\over u_2-u_\ell} - 
{qu_2-q^{-1}u_1\over u_1-u_2} \prod_{\ell=3}^k {q u_2 -q^{-1}u_\ell\over u_2-u_\ell} \right)=0 }
Hence the function $\varphi_k$ is bounded and has no pole in $u_1$, hence is independent of $u_1$,
but as it is symmetric, it is a constant, say $C_k$. To compute it, we take the limit $u_1\to \infty$, and find
the recursion relation
$\varphi_k(u_1,\ldots,u_k)=q^{k-1}+ q^{-1} \varphi_{k-1}(u_2,\ldots,u_k)$, henceforth $C_k=q^{k-1}+q^{-1}C_{k-1}$. Moreover,
by direct inspection, we find $C_1=0$, therefore the sequence $C$ is entirely fixed, and coincides with that of  the Chebyshev
polynomials of the second kind, namely $C_k=U_{k-1}=(q^k-q^{-k})/(q-q^{-1})$. This completes the proof of \antiqvander.

Applying \antiqvander\ to \expliciST, and decomposing when necessary the permutation $\sigma$ according
to the images of $1$ and/or $k$, say $\sigma(1)=j$ and $\sigma(k)=m$, we arrive at
\eqn\arriveST{\eqalign{
&{{\cal A}\Big(T(u_1,\ldots,u_k)\Big)\over (-1)^{k(k-1)/2}U_1 \cdots U_{k-1} \Delta({\bf u})} 
=\left((q z_{i+1}-q^{-1}z_{i})\prod_{\ell=1}^k f_\ell g_\ell-(qz_{i}-q^{-1}z_{i+1})\right)\cr
&-(z_{i+1}-z_i)\Bigg\{ U_{k-2}U_{k-3} -U_{k-3}\sum_{j=1}^k f_j\prod_{\scriptstyle\ell=1\atop\scriptstyle\ell\neq j}^k 
{q u_j -q^{-1}u_\ell\over u_j-u_\ell}
-U_{k-3}\sum_{m=1}^k g_m\prod_{\scriptstyle \ell=1\atop\scriptstyle\ell\neq m}^k 
{q u_\ell -q^{-1}u_m\over u_\ell-u_m} \cr
&\ \ \ \ \ \ \ \ \ \ \ \ \ \ \ \ \ \ +\sum_{1\leq j\neq m\leq k} f_j g_m {q u_j-q^{-1}u_m\over u_j-u_m}
\prod_{\scriptstyle\ell=1\atop \scriptstyle\ell\neq j,m}^k 
{q u_j-q^{-1}u_\ell\over u_j-u_\ell}{q u_\ell-q^{-1}u_m\over u_\ell-u_m}\Bigg\} 
 \cr}}
 where we use the notations \deffgs.
 Our last task is to prove that the r.h.s. of \arriveST\ vanishes identically (we denote it by $B$ in the following). 
 To this end, we view it as a rational fraction
 of the variable $z_{i+1}$, with possible poles at $z_{i+1}\to u_s$, $\ell=1,2,\ldots,k$ and at infinity. We first compute
 the residue at $z_{i+1}\to u_s$. From the definition \deffgs, we have 
 Res$_{z_{i+1}\to u_s}(g_s)=g'_s=-(q-q^{-1})u_s$, and all other
 terms have a finite limit, henceforth:
 \eqn\resius{ \eqalign{ {1\over g_s'}&{\rm Res}_{z_{i+1}\to u_s}(B)= (q u_s-q^{-1}z_i)\prod_{\ell=1}^k f_\ell 
 \prod_{\scriptstyle\ell=1\atop \scriptstyle\ell\neq s}^k 
{q u_\ell-q^{-1}u_s\over u_\ell-u_s}\cr
 &-(u_s-z_i)\left\{ \sum_{\scriptstyle j=1\atop \scriptstyle j\neq s}^k
 f_j \prod_{\scriptstyle\ell=1\atop \scriptstyle\ell\ne s, j}
 {q u_j-q^{-1}u_\ell\over u_j-u_\ell}-U_{k-3}\right\}
 \prod_{\scriptstyle\ell=1\atop \scriptstyle\ell\neq s}^k 
{q u_\ell-q^{-1}u_s\over u_\ell-u_s}\cr
 &= (u_s-z_i)\prod_{\scriptstyle\ell=1\atop \scriptstyle\ell\neq s}^k 
{q u_\ell-q^{-1}u_s\over u_\ell-u_s} \left\{  \prod_{\scriptstyle\ell=1\atop \scriptstyle\ell\neq s}^k 
f_\ell -
 \sum_{\scriptstyle j=1\atop\scriptstyle j\neq s}^k f_j 
\prod_{\scriptstyle\ell=1\atop \scriptstyle\ell\neq s,j}^k {q u_j-q^{-1}u_\ell\over u_j-u_\ell}
 +U_{k-3} \right\} \cr}}
 Noting that the last bracket involves only functions of the $k-1$ variables $u_1,\ldots,u_{s-1},u_{s+1},\ldots, u_k$, its vanishing
 is actually the consequence of the following lemma, valid for all $p\geq 1$:
 \eqn\lemafU{ \prod_{\ell=1}^p f_\ell -\sum_{j=1}^p f_j 
\prod_{\scriptstyle\ell=1\atop \scriptstyle\ell\neq j}^p {q u_j-q^{-1}u_\ell\over u_j-u_\ell} +U_{p-2}=0}
 Proof. Viewed as a function of $z_i$ via the definition \deffgs, 
 the l.h.s. of \lemafU\ (which we denote by $D$) is a rational fraction, 
 with possible poles at $z_i\to q^2 u_\ell$,
 $\ell=1,2,\ldots,p$ and at infinity. Let us first compute the residue at $z_i\to q^2u_s$, for which the only contributions
 come from Res$_{z_i\to q^2u_s}(f_s)=(q-q^{-1})u_s=f_s'$:
 \eqn\compuresidu{ {1\over f_s'}{\rm Res}_{z_i\to q^2u_s}(D)= 
 \prod_{\scriptstyle\ell=1\atop\scriptstyle\ell\neq s}^p {u_\ell-q^2u_s\over q (u_\ell-u_s)}
 -\prod_{\scriptstyle\ell=1\atop\scriptstyle\ell\neq s}^p {q u_s-q^{-1}u_\ell\over u_s-u_\ell} =0}
 So $D$ has no finite pole, and it is moreover bounded at infinity, with limit
 \eqn\limiteD{ \lim_{z_i\to \infty} (D)= q^{p} -q \sum_{j=1}^p 
\prod_{\scriptstyle\ell=1\atop\scriptstyle\ell\neq p}^p  {q u_j-q^{-1}u_\ell\over u_j-u_\ell} +U_{p-2}}
 Applying the above sublemma \sublemmauseful, we may rewrite this into
 $q^{p} -q U_{p-1}+U_{p-2}=0$. We conclude that $D=0$, and henceforth $B$ has no finite pole in $z_{i+1}$. We must now examine
 possible residues at $z_{i+1}\to \infty$. The leading behavior of $B$ when $z_{i+1}\to \infty$ is polynomial of degree $\leq 1$.
 Noting that all $\lim_{z_{i+1}\to \infty}g_\ell= q^{-1}$, 
 the coefficient of $z_{i+1}$ of $B$ in this limit reads:
 \eqn\readcoef{\eqalign{
 B\vert_{z_{i+1}} &=q^{-1}+q^{1-k}\prod_{\ell=1}^k f_\ell  -\left\{ 
 \sum_{j=1}^k q^{-1} f_j \prod_{\scriptstyle\ell=1\atop\scriptstyle\ell\ne j}^k 
{q u_j-q^{-1}u_\ell\over u_j-u_\ell}\sum_{\scriptstyle m=1\atop\scriptstyle m\neq j}^k  
 \prod_{\scriptstyle\ell=1\atop\scriptstyle\ell\neq j,m}^k{qu_\ell-q^{-1}u_m\over u_\ell-u_m} \right. \cr
 &\left. -U_{k-3} \sum_{j=1}^k\Big( f_j \prod_{\scriptstyle\ell=1\atop\scriptstyle \ell\neq j}^k 
 {qu_j-q^{-1}u_\ell\over u_j-u_\ell} + q^{-1} \prod_{\scriptstyle\ell=1\atop\scriptstyle \ell\neq j}^k 
{qu_\ell-q^{-1}u_j\over u_\ell-u_j} \Big)
 + U_{k-2}U_{k-3}\right\} \cr}}
 Applying the sublemma  \sublemmauseful\ with $k\to k-1$ in the first line and $k$ in the second, this simplifies into
 \eqn\simplicius{ B\vert_{z_{i+1}} =q^{-1}+q^{1-k}\prod_{\ell=1}^k f_\ell  -(q^{-1} U_{k-2}-U_{k-3} )
  \sum_{j=1}^k f_j \prod_{\scriptstyle\ell=1\atop\scriptstyle\ell\ne j}^k {q u_j-q^{-1}u_\ell\over u_j-u_\ell} -U_{k-3}(U_{k-2}-q^{-1}U_{k-1})}
  After noting that  $q^{-1} U_{k-2}-U_{k-3}=q^{1-k}$ and $U_{k-2}-q^{-1}U_{k-1}=-q^{-k}$, 
  we finally apply the lemma \lemafU\ with $p=k$, with the result
  \eqn\reslutfin{ B\vert_{z_{i+1}} =q^{-1}-q^{1-k} U_{k-2} +q^{-k}U_{k-3} =q^{-k}(q^{k-1}+U_{k-3}-q U_{k-2})=0 }
  Hence we conclude that $B$ has no finite pole in $z_{i+1}$ and is bounded at $z_{i+1}\to \infty$, it is therefore
  independent of $z_{i+1}$. We now evaluate $B$ at $z_{i+1}=0$, in which case all $g_\ell=q$, and
  \eqn\fineval{B\vert_{z_{i+1}= 0}=-z_i\left(q+q^{k-1}\prod_{\ell=1}^k f_\ell -(q U_{k-2} -U_{k-3})\sum_{j=1}^k f_j 
  \prod_{\scriptstyle\ell=1\atop\scriptstyle\ell\neq j}^k {q u_j-q^{-1}u_\ell\over u_j-u_\ell} -U_{k-3}(U_{k-2}-q U_{k-1})\right)}
  where we have used again the sublemma \sublemmauseful.
  Again, we note that $q U_{k-2} -U_{k-3}=q^{k-1}$ and $U_{k-2}-q U_{k-1}=-q^k$, and we apply the lemma
  \lemafU\ with $p=k$ to get
  \eqn\lutfin{ B\vert_{z_{i+1}= 0}=-z_i( q-q^{k-1} U_{k-2} +q^k U_{k-3})= -z_i q^k(q^{1-k}+U_{k-3}- q^{-1}U_{k-2})=0 }
  We may therefore conclude that $B=0$ identically, which implies that ${\cal A}(T)=0={\cal A}(S)$, 
  which in turn implies \toprove, as explained above.
  This completes the proof of Theorem 2.

\subsec{Case of reflecting boundary conditions}
All that has been described above can for example apply to the case of periodic boundary 
conditions treated in \ZJDF, avoiding
the lengthy discussion found in this paper; in this case the function $F$ is just $1$.
In the present case of reflecting boundaries,
the solution to the $q$KZ equation must incorporate the new boundary conditions, cf Eqs.~\qkz{b,c}, which 
correspond to a non-trivial
choice of $F$, namely:
\eqnn\openpsi
$$\eqalignno{ \Psi&_{a_1,\ldots,a_n}(z_1,\ldots,z_N)=
\prod_{1\leq i<j\leq N} (q z_i-q^{-1}z_j)(q^4-z_iz_j)
\oint\cdots\oint \prod_{m=1}^n  
{d w_\ell\over 2\pi i}\cr
&
{\prod_{1\leq \ell<m\leq n} (w_m-w_\ell)(q w_\ell-q^{-1}w_m)(q^2-w_\ell w_m) 
\prod_{1\leq \ell\leq m\leq n} (q^4-w_\ell w_m)
\over \prod_{\ell=1}^n\prod_{i=1}^{N}(q^4-w_\ell z_i)
\prod_{i=1}^{a_\ell}(w_\ell-z_i)\prod_{i=a_\ell+1}^{N}
( q w_\ell -q^{-1}z_i)}
&\openpsi
\cr}
$$
where we recall that the contours of integration encircle the $z_i$ but not $q^{-2}z_i$, nor $q^4 z_i^{-1}$.
By a computation of residues similar to what is performed in \DFZJ, it is easy to show that 
$\Psi_{a_1,\ldots,a_n}$ is in fact a polynomial in the variables $z_1,\ldots,z_N$ of degree $3n(n-1)$.

We now want to show that these $\Psi_{\bf a}$ identify with the components in the intermediate basis
of the solution of the $q$KZ equation with reflecting boundary conditions discussed in 
Sect.~3.1. We note that both quantities satisfy
the main equation \toprove\ (exchange relation) of the $q$KZ system, as a consequence of 
Theorems 1 and 2 above. 
Furthermore, direct computation shows that when $a_i=i$ for $i=1,2,...,n$, the only 
non-vanishing contribution to the integral \openpsi\ comes from the multiresidue at $w_i\to z_i$, for $i=1,2,...,n$.
Cancelling all denominators, we are left with the polynomial:
\eqn\initialvalue{
\Psi_{1,2,\ldots,n}=\prod_{1\le i<j\le n} (q\,z_i-q^{-1}z_j)(q^2-z_iz_j)
\prod_{n+1\le i<j\le 2n} (q\,z_i-q^{-1}z_j)(q^4-z_iz_j)}
which is nothing but $\Psi_{\pi_0}$ \base,
and indeed the change of basis implies that $\Psi_{1,2,\ldots,n}=\Psi_{\pi_0}$. Moreover, introducing
the lexicographic order on non-decreasing sequences $(a_1,\ldots,a_n)$, we note that the sequence
$(1,2,...,n)$ is the smallest yielding a non-zero result for the integral \openpsi. Indeed, for any strictly
smaller sequence, at least two residues will have to be taken at identical points say
$w_\ell,w_m\to z_i$, causing the result to vanish, due to the Vandermonde determinant of the $w$'s
in the numerator.

The final step is to show that every component $\Psi_{\bf a}$ can be deduced from 
$\Psi_{1,2,\ldots,n}$ by use of Eq.~\toprove.
With respect to the lexicographic order on the non-decreasing sequences $(a_1,\ldots,a_n)$, 
Eq.~\toprove\ can be considered as a triangular linear system, allowing to compute
$\Psi_{a_1,\ldots,a_{m-1},\underbrace{\scriptstyle i,\ldots,i}_{k-1},i+1,a_{m+k},\ldots,a_n}$ in terms of other $\Psi_{\bf a}$
with smaller index. It is easy to conclude from this that the $\Psi_{\bf a}$ are entirely determined by the non-zero component with smallest index \initialvalue.

\subsec{Homogeneous limit}
Let us now evaluate the homogeneous limit of the components of $\Psi$, by setting
$z_1=z_2=\cdots=z_{2n}=1$, and by renormalizing it in such a way that $\Psi_{\pi_0}=1$.
This amounts to taking the integral formula \openpsi, substituting $z_i=1$ for all $i$,
and dividing it out by $\Psi_{\pi_0}(1,1,...,1)=q^{3n(n-1)/2} (q-q^{-1})^{2n(n-1)} (q+q^{-1})^{n(n-1)/2}$.

Performing then the  change of variables $w_i={1-q u_i\over 1-q^{-1}u_i}$
in the resulting expresion, we get:
$$\eqalign{J_{a_1,\ldots,a_n}(\tau)&= \tau^{n(2n-1)}
\oint\cdots\oint \prod_{m=1}^n {du_m\over 2\pi i u_m^{a_m}} \left[ 
{ \prod_{1\leq \ell \leq m \leq n}
(\tau+(\tau^2-1) (u_\ell+u_m)+\tau(\tau^2-2) u_\ell u_m)\over \prod_{m=1}^n (\tau+(\tau^2-1) u_m)^{2n}} \right. \cr
&\times\left.  \prod_{1\leq \ell<m \leq n} (u_m-u_\ell) (1+ \tau u_m+ u_\ell u_m)
(1+\tau (u_\ell+u_m )+(\tau^2-1) u_\ell u_m) \right]\cr}
$$
Note that the integration contours encircle $0$ but not $U_2/U_1=(1-\tau^2)/\tau$.
This integral formula is quite cumbersome and proves quite difficult to deal with. Instead,
let us use the inversion property \inversion\ to express the homogeneous components as the
reflected components $\Psi_{\rho(\pi)}$ at $z_1=z_2=\cdots=z_{2n}=q^3$. This simplifies drastically
the pole structure of \openpsi, and finally, after division by $\Psi_{\pi_0}(1,1,...,1)$, leads to the following
integrals:
\eqnn\clopsi
$$\eqalignno{ K_{b_1,\ldots,b_n}(\tau) &= \oint\cdots\oint \prod_{\ell=1}^n {du_\ell\over 2\pi i u_\ell^{b_\ell}} \left[ 
\prod_{1\leq \ell \leq m \leq n} (1-u_\ell u_m) \right.\cr
&\times \left. \prod_{1\leq \ell < m \leq n} (u_m-u_\ell)
(1+\tau u_m+u_\ell u_m) (\tau+u_\ell+u_m) \right]&\clopsi \cr}
$$
obtained by performing the change of variables $w_i=q{1-q u_i\over 1-q^{-1} u_i}$. We have the relation 
$J_{{\bf a}(\pi)}=K_{{\bf b}(\pi)}$ for all $\pi\in LP_{2n}$, hence
we may call the $K$'s  the ``arch closing'' basis elements.

The expression in brackets in formula~\clopsi\ can be be thought of as 
a (polynomial) ``quasi-generating function'' for the $K_{\bf b}$, that is the homogeneous limit of the 
$\Psi_{\bf a}/\Psi_{1,2...,n}$.
An important consequence of this formula is that the $K_{\bf b}$ 
are polynomials with integer coefficients in $\tau$.
Since the change of basis from the intermediate basis to the basis of link patterns has entries that are also polynomials
with integer coefficients in $\tau$, we conclude that the homogeneous $\Psi_\pi(\tau)$,
normalized by $\Psi_{\pi_0}(\tau)=1$, possess the same property,
as had been conjectured in \DFVS. In particular, at $\tau=1$, where the $\Psi_\pi(1)$ are identified with the ground state components of the loop model with reflecting boundaries normalized by $\Psi_{\pi_0}(1)=1$, we conclude
that all these components are integers, as had been conjectured earlier in \OPRS.

Note that the same integrality argument applies equally well to the case of periodic 
boundary conditions treated in \ZJDF.

\newsec{ Even case: proofs of various conjectures}
\subsec{Solution at $\tau\to 0$}
When $\tau\to 0$, the integral \clopsi\ is easily evaluated upon changing to variables $v_m=u_m/\tau$
and explicitly retaining only the leading terms when $\tau\to 0$ in each factor. This results in
\eqn\leftau{ K_{b_1,\ldots,b_n}(\tau)\sim \tau^{n^2-\sum b_i} 
\oint\cdots\oint\prod_{m=1}^n  {dv_m\over 2\pi i v_m^{b_m}}
\prod_{1\leq \ell<m \leq n}(v_m-v_\ell)(1+v_\ell+v_m )
}
Upon identifying the product as the Vandermonde determinant $ \Delta(v(1+v))$ (using the notation 
$\Delta(z)\equiv \prod_{1\leq \ell<m\leq n} (z_m-z_\ell)$ for the Vandermonde determinant
 of $z\equiv \{z_1,\ldots,z_n\}$, also equal to $\det(z_\ell^{m-1})_{\ell,m=1,2,\ldots,n}$), we may recast \leftau\
 into a single determinant:%
\eqnn\singdet
$$\eqalignno{ K_{b_1,\ldots,b_n}(\tau) &\sim \tau^{n^2-\sum b_i} \det_{1\leq \ell,m \leq n}\left(
 \oint {dv \over 2\pi i v} v^{m-b_\ell} (1+v)^{m-1} \right) \cr
 &\sim    \tau^{n^2-\sum b_i} \det_{1\leq \ell,m\leq n}
 {m-1\choose b_\ell -m}  &\singdet \cr}
$$
 The latter determinant is nothing but the number $N_{10}(b_1,b_2,\ldots,b_n)$ of NILP in bijection
with TSSCPP with fixed
 endpoints $b_1,b_2,\ldots,b_n$ in their NILP formulation, while the power of $\tau$ reads
 $n(n-1)-\sum (b_i-1)=\beta(\pi)$, which is also the number of boxes in the box decomposition
 of the Dyck path associated to $\pi$. 
 
 This result may now be immediately translated into an estimate for the small $\tau$ behavior
 of the $q$KZ solution in the link pattern basis. Indeed, the change of basis \matrirela\ allows to
 identify $\Psi_\pi(\tau) \sim K_{{\bf b}(\pi)}(\tau)$ when $\tau\to 0$. This is readily seen by writing
 \eqn\invchgbas{ \Psi_\pi(\tau)=\sum_\alpha C^{-1}(\tau)_{\pi,\alpha} K_{{\bf b}(\alpha)}(\tau)}
 and recalling that $C^{-1}(\tau)$ has all entries polynomial in $\tau$, and that it is lower triangular
 with respect to containment order of the Dyck paths associated to the link patterns, we deduce
 that any $\pi$ such that $C^{-1}(\tau)_{\pi,\alpha}$ is non-zero must be contained in $\alpha$, hence have
 a strictly smaller number of boxes if it is distinct from $\alpha$. As $ K_{{\bf b}(\alpha)}(\tau)$ behaves
 like $\tau^{n(n-1)-\beta(\alpha)}$, we deduce that any contribution to the sum \invchgbas\ with
 $\alpha\neq \pi$ is subleading, as $\beta(\alpha)<\beta(\pi)$.
 
 This completes the proof of the small $\tau$ conjecture of Ref.~\DFVS, in the
 case of even size, namely that
 \eqn\conjpsieven{ \Psi_{\pi}(\tau)\sim \tau^{\beta(\pi)} \ N_{10}\Big(b_1(\pi),\ldots,b_n(\pi) \Big) }
 where $b_i(\pi)$ denote the positions of the arch closures of $\pi$, counted from right to left.

\subsec{Solution at large $\tau$}
For large $\tau$, we obtain the leading contribution to $K_{b_1,\ldots,b_n}(\tau)$ by changing variables to 
$v_m=\tau u_m$ in the integral formula \clopsi, and retaining only the leading order in $\tau$ within each factor in the integrand. This yields
\eqn\leadtau{ K_{b_1,\ldots,b_n}(\tau)\sim \tau^{\sum (b_i -1)} 
\oint\cdots\oint \prod_{\ell=1}^n {dv_\ell\over 2\pi i v_\ell^{b_\ell}} 
\Delta(v) \prod_{\ell=1}^{n} (1+v_\ell)^{\ell-1} }
By multilinearity of the Vandermonde determinant, this may be recast into%
\eqnn\recast
$$\eqalignno{
K_{b_1,\ldots,b_n}(\tau)&\sim \tau^{\sum (b_i -1)} 
\det_{1\leq \ell,m\leq n}\left( \oint{dv \over 2\pi i v} v^{m-b_\ell}(1+v)^{\ell-1}\right) \cr
&\sim \tau^{\sum  (b_i -1)} \det_{1\leq \ell,m\leq n} {\ell-1 \choose b_\ell -m} &\recast\cr}
$$

To translate this into a result for $\Psi_\pi(\tau)$, let us again consider the change of basis 
\invchgbas, and note that  $\sum (b_i(\pi)-1)=n(n-1)-\beta(\pi)$. We wish to prove that
$\Psi_\pi(\tau)\sim K_{{\bf b}(\pi)}(\tau)$ at large $\tau$.

\fig{The strip decomposition of a typical Dyck path. To each ascending step $(i-1,h-1)\to (i,h)$
we associate a diagonal row of $h-1$ boxes as indicated, thus forming a strip of boxes.
Strips associated to ascending steps exhaust all the boxes in the decomposition of the Dyck path.}{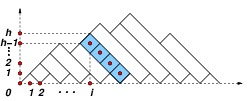}{8.cm}
\figlabel\stripdec
The degree of the matrix element $C_{\alpha,\pi}(\tau)$ in $\tau$ is given by the following
quantity. Define first $h(\pi,\alpha)$ as the sum over the arch openings of $\alpha$
of the total number of arches of $\pi$ sitting above their position (an arch $(i,\pi(i))$ is said to sit
above position $j$ iff $i\leq j<\pi(i)$). The quantity $h(\pi,\alpha)$ is also the sum
over the heights $h_i(\pi)$ in the Dyck path of $\pi$ (or equivalently the position occupied by the path
on the integer half-line at time $i$),
measured at the positions $i$ of the points in the Dyck path of $\alpha$
reached by an ascending step (i.e. such that $h_i(\alpha)=h_{i-1}(\alpha)+1$), namely:
\eqn\defhalpi{ h(\pi,\alpha)=\sum_{\scriptstyle i=1\atop\scriptstyle h_i(\alpha)=h_{i-1}(\alpha)+1}^{2n} h_i(\pi)}
With this expression, it is easy to see that $h(\pi,\pi)=\beta(\pi)+n$
for any $\pi\in LP_{2n}$. Indeed, $h(\pi,\pi)$ is the sum of heights of ends of ascending steps in 
the Dyck path of $\pi$.
As illustrated in Fig.\stripdec, we may associate to each such ascending step 
$(i-1,h_{i-1}(\pi))\to(i,h_{i}(\pi))$ with $h_i(\pi)=h_{i-1}(\pi)+1$
the diagonal strip of $h_i(\pi)-1$ boxes with centers at $(i+\ell-1,h_i(\pi)-\ell)$, $l=1,2,...,h_i(\pi)-1$,
and this exhausts all boxes of $\pi$. Such a ``strip decomposition'' was considered in Ref.~\TLAMEAN.
We deduce that $\beta(\pi)=\sum_{i: h_i(\pi)=h_{i-1}(\pi)+1}(h_i(\pi)-1)=h(\pi,\pi)-n$ as there are exactly
$n$ ascending and $n$ descending steps in the Dyck path.
Then, using the definition  \formulA\ and the fact that the Chebyshev polynomials $U_m$
have degree $m$ in $\tau$, we have
\eqn\degA{d_{\alpha,\pi}\equiv \deg\Big(C_{\alpha,\pi}(\tau)\Big)=h(\pi,\alpha)-h(\pi,\pi) }

Finally we need the following lemma:

{\sl The quantity $f_\pi(\alpha)=h(\pi,\alpha)+h(\alpha,\alpha)$, where $\alpha$ runs over
the link patterns whose Dyck path is included in that of $\pi$, reaches its 
maximum at $\alpha=\pi$ only.}

\par\noindent Proof: let us show that $f_\pi(\alpha)$ is a non-decreasing function with the size of $\alpha$,
namely the number of boxes in the decomposition of its Dyck path. Assume $\alpha'\in LP_{2n}$
differs from $\alpha$ by a single box say at positions $i-1,i,i+1$ in the Dyck path formulation, with
identical heights $h_j(\alpha')=h_j(\alpha)+2\delta_{j,i}$ except at the position $i$.
We see easily that $h(\alpha',\alpha')=h(\alpha,\alpha)+1$ as the ascending step
$(i,i+1)$ in $\alpha$ is replaced by $(i-1,i)$ in $\alpha'$, and $h_i(\alpha')=h_{i+1}(\alpha)+1$. 
Moreover, 
\eqn\compah{ h(\pi,\alpha')=h(\pi,\alpha)+h_i(\pi)-h_{i+1}(\pi) \geq h(\pi,\alpha)-1}
as $h_i(\pi)\geq h_{i+1}(\pi)-1$. We deduce
that $f_\pi(\alpha')\geq f_\pi(\alpha)$, hence that $f_\pi(\pi)\geq f_\pi(\alpha)$ for all
$\alpha$ whose Dyck path is included in that of $\pi$. Finally, $\pi$ is the unique maximum of
$f_\pi(\alpha)$, as is easily seen by removing a box from $\pi$ say at position $i$
and comparing heights in the Dyck paths. Indeed, we have say $(h_{i-1}(\pi),h_i(\pi),h_{i+1}(\pi))=(m,m+1,m)$,
and in the box-removed $\pi'$ we have $(h_{i-1}(\pi'),h_i(\pi'),h_{i+1}(\pi'))=(m,m-1,m)$. Hence
$h(\pi,\pi')=h(\pi,\pi)-(m+1)+m$ and $h(\pi',\pi')=h(\pi,\pi)-(m+1)+m$, so that $f_\pi(\pi')=f_\pi(\pi)-2$.

We deduce from the above lemma an upper bound on the degree $d_{\alpha,\pi}$  \degA:
$d_{\alpha,\pi}=h(\pi,\alpha)-h(\pi,\pi)< h(\pi,\pi)-h(\alpha,\alpha)=\beta(\pi)-\beta(\alpha)$.
We may now prove that the large $\tau$ contribution to $\Psi_\pi(\tau)$ is given by that
to $K_{{\bf b}(\pi)}(\tau)$. This is done by induction on the (decreasing) number of boxes in $\alpha$. 
Assume it is true for all $\pi$ with a strictly larger number of boxes than $\alpha$. Then for all these $\pi$,
$\Psi_\pi(\tau)$ has degree equal to that of $K_{{\bf b}(\pi)}(\tau)$, namely $n(n-1)-\beta(\pi)$.
Then the expression $C_{\alpha,\pi}(\tau)\Psi_\pi(\tau)$, for $\alpha\subset \pi$ and $\pi\neq \alpha$
has degree $d_{\alpha,\pi}+n(n-1)-\beta(\pi)<n(n-1)-\beta(\alpha)$ by the above inequality.
But $K_{{\bf b}(\alpha)}(\tau)=\sum_{\pi} C_{\alpha,\pi}(\tau)\Psi_\pi(\tau)$ has degree
$n(n-1)-\beta(\alpha)$, hence it must be attained by the term $\pi=\alpha$ in the sum, and
we deduce that $\Psi_\alpha(\tau)\sim K_{{\bf b}(\alpha)}(\tau)$ for large $\tau$. 
As the result holds trivially for the largest link pattern $\pi_0$ (the only non-vanishing matrix element
of $C$ with this first entry is just $C_{\pi_0,\pi_0}(\tau)=1$),
this completes the desired proof.

\subsec{Generalized sum rule}
The fundamental remark of Ref.~\ZJDF\ for the even case was that summing $\Psi_{\bf a}$ over a specific subset 
of ``opening arch'' basis elements, namely the set of arch openings $a$'s such that $a_i=2i-1-\epsilon_i$,
$\epsilon_i\in \{0,1\}$, amounted to summing $\Psi_\pi$ over the whole set $LP_{2n}$. This was readily seen 
as a property of the change of basis $C_{\alpha,\pi}(\tau)$ \formulA. Due to a reflection 
symmetry property, an analogous statement may be derived for the ``arch closing'' basis elements. As a result, 
the above sum rule for $\sum_{\pi\in LP_{2n}} \Psi_\pi$ is obtained by 
summing the integrals $K_{b_1,b_2,\ldots,b_n}(\tau)$ over the $b_i=2i-1-\epsilon_i$, with $\epsilon_i\in \{0,1\}$.

More precisely, let us consider
\eqn\consid{K(t \vert \tau)\equiv \sum_{\epsilon_i\in\{0,1\}} t^{\Sigma \epsilon_i} 
K_{1-\epsilon_1,3-\epsilon_2,\ldots,2n-1-\epsilon_n}(\tau)}
This quantity has many interesting specializations: (i) $t=0$ corresponding to the ``maximum'' component
$K_{1,3,\ldots,2n-1}(\tau)=\Psi_{\pi_{max}}(\tau)$, where $\pi_{max}$ connects the points $(2i-1,2i)$ via little arches only
(ii) $t\to \infty$ corresponding to $K_{1,2,4,\ldots,2n-2}(\tau)$ which is also equal to a single
component, namely the link pattern $\pi'_{max}$ connecting $(2i,2i+1)$, and $(1,2n)$;  
(iii) $t=1$ corresponding to the sum rule
$\sum_{\pi \in LP_{2n}} \Psi_\pi= K(1|\tau)$. From the definition \clopsi, it is readily computed as%
\eqnn\redili
$$\eqalignno{ K(t\vert \tau)&=  
\oint\cdots\oint \prod_{m=1}^n
{du_m(1+t u_m)\over 2\pi i u_m^{2m-1}} \left[
\prod_{1\leq \ell \leq m \leq n} (1-u_\ell u_m) \right.\cr
&\times \left. \prod_{1\leq \ell < m \leq n} (u_m-u_\ell)
(1+\tau u_m+u_\ell u_m) (\tau+u_\ell+u_m) \right] &\redili\cr}
$$
We now make use of the following lemma, first conjectured in \ZJDF\ and then proved by Zeilberger in \Zeil.%
\eqnn\lemma
$$\eqalignno{& \left\{ \prod_{1\leq \ell \leq m \leq n} (1-u_\ell u_m)
{\cal A}\left(\prod_{m=1}^n u_m^{2-2m} \prod_{1\leq \ell < m \leq n} (1+\tau u_m+u_\ell u_m)\right) \right\}_{<0}\cr
&= \prod_{1\leq \ell < m \leq n} (u_m^{-1}-u_\ell^{-1})(\tau+u_\ell^{-1}+u_m^{-1})=\Delta(u^{-1}(\tau+u^{-1}))&\lemma\cr}
$$
where the subscript $<0$ means that we retain in the corresponding multiple Laurent series of the $u_m$ 
the terms with only non-positive powers, and the symbol ${\cal A}$ means as before antisymmetrization with 
respect to the permutations of variables: ${\cal A}(f)(u_1,\ldots,u_n)\equiv \sum_{\sigma\in S_n}(-1)^\sigma
f(u_{\sigma(1)},\ldots,u_{\sigma(n)})$. Noting that $\Delta(u)={\cal A}(\prod_{m=1}^n u_m^{m-1})$, and that 
$\oint {\cal A}(f) g=\oint f {\cal A}(g)$, and applying the lemma to \redili, we get%
\eqnn\redilib
$$\eqalignno{ 
K(t\vert \tau)&=  
\oint\cdots\oint \prod_{m=1}^n  {du_m(1+t u_m)u_m^{m-1}\over 2\pi i u_m}
\prod_{1\leq \ell < m \leq n}\!(\tau+u_\ell+u_m)\ 
{\cal A}\Big(\prod_{m=1}^n u_m^{1-m}(\tau+u_m^{-1})^{m-1}\Big)\cr
&=\oint\cdots\oint  \prod_{m=1}^n 
{{du_m\over 2\pi i u_m}(1+t u_m)u_m^{1-m}(\tau+u_m^{-1})^{m-1}} \Delta(u(\tau+u))\cr
&=\det_{1\leq \ell,m\leq n}  \left( \oint {du \over 2\pi i u} (1+t u)u^{\ell-m} (\tau+u^{-1})^{m-1}(\tau+u)^{\ell-1} \right) &\redilib\cr}
$$
where in the last step we have used the multilinearity of the Vandermonde determinant to rewrite the whole
multiple integral as a determinant of single integrals. This finally yields%%
\eqnn\finK
$$\eqalignno{ K(t\vert \tau)&= \det_{1\leq \ell,m\leq n} \left( f_{\ell,m}(t\vert \tau)\right) \cr
f_{\ell,m}(t\vert \tau)&= \sum_r \tau^{2\ell+2m-3-2r}  {\ell-1\choose r-\ell}\left(\tau {m-1\choose r-m}
+t {m-1\choose r+1-m}\right)&\finK\cr
}$$

In order to reconnect with the results of \DFVS, it is convenient to rewrite 
$K(t\vert \tau)$, by shifting all indices: $\ell=j+1$, $m=i+1$, $r=s+1$, and noting that the
first row/column do not contribute to the determinant:
\eqn\Kagain{
K(t\vert \tau)=\det_{1\leq i,j\leq n-1} \left[ \sum_s \tau^{2i+2j-2s}
{i\choose 2i-s} \left( t \tau {j\choose 2j-s+1}+{j\choose 2j-s}\right)\right]
}

\fig{(a) The fundamental domain of a modified CSTCPP and its NILP description (blue and black paths)
and (b) the correspondence (heights of the box piles) to triangular arrays of integers (c).
Colors are related to weights: red means a weight of $\tau$, purple $1/t$, green 
$x$. With these conventions, (a) receives a weight $\tau^3\times t^{-2}\times \tau^2$
(times a global factor of $t^{n-1}=t^3$),
while (b) and (c) have the weight $x^2$.}{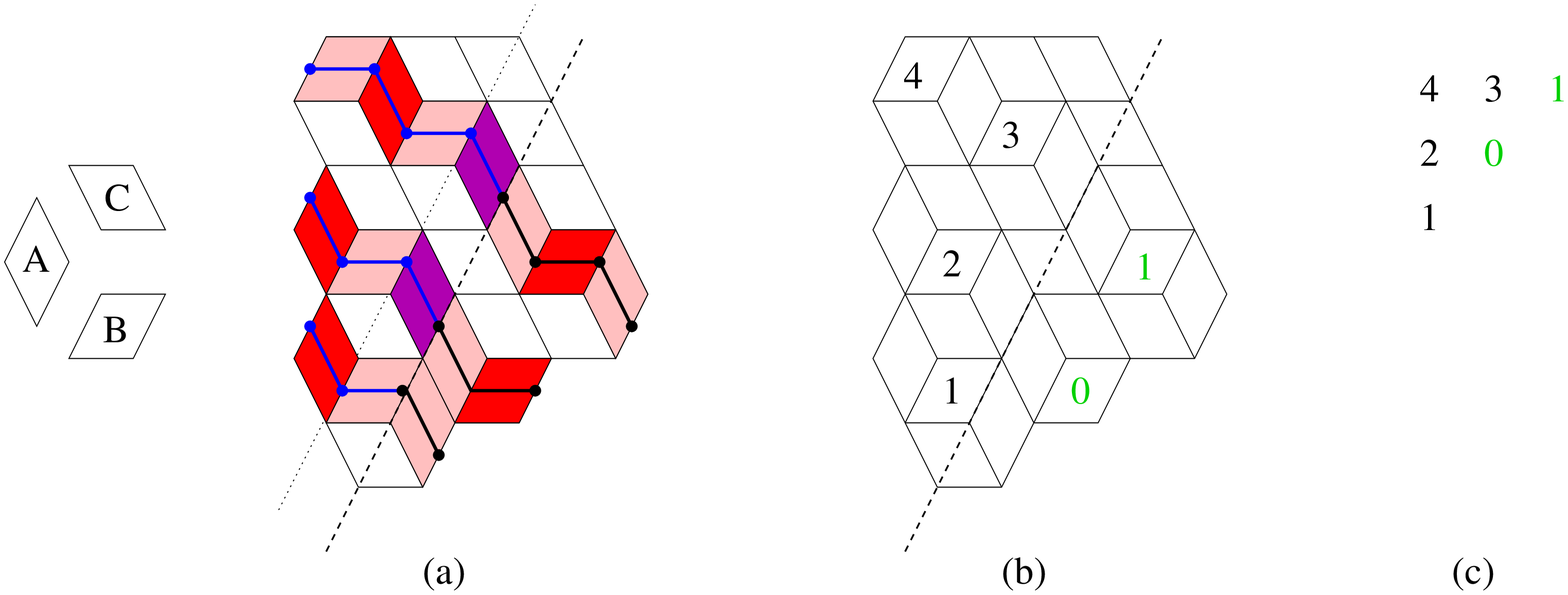}{12cm}\figlabel\cstcpptriarray
For generic $t$, the polynomials $K(t\vert \tau)$ correspond to a refined $\tau,t$-enumeration of 
CSTCPP$^\triangle$, which can be described as follows.
In the NILP formulation of the CSTCPP$^\triangle$
the CSTCPP$^\triangle$ are described by paths made by two orientations
of lozenges in a fundamental domain of the CSTCPP$^\triangle$: these are the (colored) lozenges
of types A, B on Fig.~\cstcpptriarray (a).
In  Ref.~\DFVS, these paths were viewed as pairs of paths sharing their arrival point,
the paths on one side being one step longer than on the other side:
the set of paths below the dashed line of Fig.~\cstcpptriarray (a)
was identified as the NILP in bijection with TSSCPP (paths represented in black), while that
above was viewed as an augmented one, with one more last step in each path
(paths represented in blue, with the last step taking place in the strip just above
the dashed line).

At each step of the paths, lozenges of type A above the diagonal dashed line and B below
(in red on the figure) are given a
weight of $\tau$ in $K(t\vert\tau)$,
{\it except}\/ the last step of the longer paths (in the strip just above the dashed line), 
where a factor $t$ is given to lozenges of type B; it is however more convenient
to consider for this last step 
that the factor of $\tau$ is replaced by $1/t$ (purple lozenges on the figure) up to
global multiplication by $t^{n-1}$.

Let us now discuss various specializations of $K(t\vert\tau)$.

As announced above, at $t=0$, we find the maximal component 
\eqn\maxcompeven{\Psi_{\pi_{max}}(\tau)=K(0\vert \tau)= \det_{1\leq i,j\leq n-1} \left[
\sum_s \tau^{2i+2j-2s} {i\choose 2i-s}{j\choose 2j-s} \right] }
This matches the expression conjectured in \DFVS\ (conjecture 3, Eq.~(5.1)).
Likewise, at $t\to \infty$, we find:
\eqn\rotcomp{ \Psi_{\pi_{max}'}(\tau)=\lim_{t\to\infty}{1\over t^n}K(t\vert \tau)=
\det_{1\leq i,j\leq n-1} \left[ 
\sum_s \tau^{2i+2j-2s+1}  {i\choose 2i-s}{j\choose 2j-s+1} \right]}

We also obtain at $t=1$ the sum rule for the components of $\Psi$:
\eqn\sumruleven{
\sum_{\pi\in LP_{2n}} \Psi_\pi(\tau)=  
\det_{1\leq i,j\leq n-1} \left[ \sum_s \tau^{2i+2j-2s}
{i\choose 2i-s} \left( \tau {j\choose 2j-s+1}+{j\choose 2j-s}\right)\right]
}
which matches the conjectured expression found in Ref.~\DFVS\ (Eq.~(4.10)).
%\eqn\identif{\eqalign{ \tau^{2\ell+2m-2-2r}  {\ell-1\choose r-\ell}{m-1\choose r-m} 
%&=\tau^{2i+2j-2s}{i\choose 2i-s}{j\choose 2j-s} \cr
%\tau^{2\ell+2m-3-2r}  {\ell-1\choose r-\ell}{m-1\choose r+1-m} 
%&=\tau^{2i+2j-2s+1}{i\choose 2i-s}{j\choose 2j-s+1} \cr}}

Finally, one more identification is of some interest. 
Setting $t=\tau^{-1}$ we find the generating function $T_n(x=\tau^2,1)$ of \Rob.
The latter has several intepretations. One of them is the following:
consider triangular arrays of non-negative integers $a_{ij}$, $i,j\ge 1$, $i+j\le n$,
with weakly decreasing rows and columns and such that
$a_{i1}\le n-i+1$ for all $i$.
These arrays turn out to be in bijection with 
CSTCPP$^\triangle$, see Fig.~\cstcpptriarray(b); to produce $T_n(x,1)$ one 
gives a weight $x$ to parts $a_{ij}$ such that $a_{ij}\le j-1$, see Fig.~\cstcpptriarray(c).
Via the bijection 
this corresponds in terms of plane partitions to lozenges of type B
that are below the diagonal.
We now show that this is the same weight that
is given to CSTCPP$^\triangle$ in $K(\tau^{-1}|\tau)$. When $t=\tau^{-1}$ all red/purple lozenges
get a weight of $\tau$. Call $n_{ab}$ the number of lozenges of type $a$ in region $b$
where $a=A,B,C$ and $b=\uparrow,\downarrow$ corresponds to above/below the diagonal 
dashed line.
Then the weight for CSTCPP$^\triangle$ is $\tau^{n_{A\uparrow}+n_{B\downarrow}}$, whereas
the weight for triangular arrays is $x^{n_{B\downarrow}}$.
Now the number of tiles of each orientation is fixed (independent
of the choice of plane partition): in particular $n_{A\uparrow}+n_{A\downarrow}=n(n+1)/2-1$.
Furthermore the number of tiles of the first 2 types
in each region is also fixed: $n_{A\uparrow}+n_{B\uparrow}=n(n-1)/2$.
We conclude immediately that $n_{A\uparrow}+n_{B\downarrow}=2n_{B\downarrow}+n-1$, which allows us to identify
the weights taking into account the prefactor $t^{n-1}$ and the equality $x=\tau^2$.

$T_n(x,1)$ is also conjectured to be the $x$-enumeration of VSASM where a weight $x$ is given 
to each pair of $-1$s (conj.~3.2 of \Rob). In the current framework there is no obvious explanation
of this coincidence.

\newsec{The odd case}
\subsec{General solution}
The link patterns in size $2n+1$ are obtained bijectively from those in size one more $2n+2$, by
simply erasing the rightmost arch and suppressing the rightmost endpoint $2n+2$, so that the 
opening point of the rightmost arch remains isolated and unmatched. 
The component of $\Psi$ in size $2n+1$
can then be obtained from the corresponding one in size $2n+2$ by setting $z_N=0$.
There is of course another bijection which consists on the contrary in erasing the leftmost arch and
endpoint, so that the closing point of the leftmost arch remains unmatched. This time setting $z_1=\infty$
allows to obtain $\Psi$ in odd size $2n+1$ from even size $2n+2$. Since the $q$KZ equation is not quite
left-right symmetric these two routes lead to different expressions.

Here we use the second route, having in mind that eventually we shall apply left-right symmetry as before
to express everything in terms of closings instead of openings.
We start with Eq.~\openpsi\ at $N=2n+2$ with $a_1=1$ and integrate over $w_1$;
we set $z_1=\infty$ at this stage and obtain after shifting one step the indices of the
$a$'s, $w$'s and $z$'s:
\eqnn\openpsiodd
$$\eqalignno{ \Psi'&_{a_1,\ldots,a_n}(z_1,\ldots,z_N)=
\prod_{1\leq i<j\leq N} (q z_i-q^{-1}z_j)(q^4-z_iz_j)
\oint\cdots\oint \prod_{m=1}^n  
{d w_\ell\over 2\pi i}\cr
&
{\prod_{1\leq \ell<m\leq n} (w_m-w_\ell)(q w_\ell-q^{-1}w_m)(q^2-w_\ell w_m) 
\prod_{1\leq \ell\leq m\leq n} (q^4-w_\ell w_m)
\over \prod_{\ell=1}^n\prod_{i=1}^{N}(q^4-w_\ell z_i)
\prod_{i=1}^{a_\ell}(w_\ell-z_i)\prod_{i=a_\ell+1}^{N}
( q w_\ell -q^{-1}z_i)}
&\openpsiodd
\cr}
$$
that is, exactly the same expression as Eq.~\openpsi, but in which now $N=2n+1$.
It provides us with $\Psi$ in odd size in terms of the openings $a_i$, with the convention that the
closing of the former leftmost arch is unmatched (the opening, i.e.\ the leftmost point, being erased).

We now use left-right symmetry in the form of the inversion property \inversion, to obtain the 
component $\Psi'_{b_1,\ldots,b_n}$ defined
in terms of the $n$ closings $b_i$ counted from right to left, with the convention that 
the opening of the former rightmost arch
is unmatched. We simply take Eq.~\openpsiodd\ and substitute $a_i$ with $b_i$ and $z_i$ 
with $q^3/z_i$. Note that $b_i\leq 2i-1$ for closings to the right of the unmatched point
and $b_i\leq 2i$ for closings to its left.
We can now take the homogeneous limit $z_i=1$ and obtain
\eqnn\clodd
$$\eqalignno{
K'_{b_1,b_2,\ldots,b_n}(\tau)&=
\oint\cdots\oint\prod_{m=1}^n
{du_m(1+\tau u_m+u_m^2)\over 2\pi i u_m^{b_m}} \Bigg[
\prod_{1\leq \ell \leq m \leq n} (1-u_\ell u_m) \cr
&\times \prod_{1\leq \ell < m \leq n} (u_m-u_\ell)
(1+\tau u_m+u_\ell u_m) (\tau+u_\ell+u_m) \Bigg] &\clodd\cr}
$$

\subsec{Solutions at $\tau\to 0$ and large $\tau$}
Repeating the calculations of Sects.~5.1 and 5.2, and using the same changes of variables,
we arrive easily at the following formulas, respectively for $\tau\to 0$:
\eqn\zerodd{K'_{b_1,\ldots,b_n}(\tau)\sim \tau^{n(n-1)-\sum (b_i-1)} \det_{1\leq \ell,m\leq n}
 {m-1\choose b_\ell -m}}
and $\tau\to\infty$:
\eqn\infiodd{K'_{b_1,\ldots,b_n}(\tau)\sim \tau^{\sum (b_i-1)}  \det_{1\leq \ell,m\leq n}
{\ell \choose b_\ell -m} }

Note that the determinants \zerodd\ vanish when some of the $b_j$'s attains $2j$, hence the correct 
small $\tau$ behaviour is actually subleading in those cases. The answer \zerodd\ is accurate
only if the unmatched point of the corresponding link pattern is at position $1$
(in which case all $b_i\leq 2i-1$). In general,
the correct answer depends on the position of this point, and requires a further 
expansion in powers of $\tau$, that we shall omit here. 

Note also that the size $2n+1$  large $\tau$ solution \infiodd\ is related to 
the one in size $2n+2$ \recast\ via:
\eqn\linkodeven{ K'_{b_1,\ldots,b_n}(\tau)\sim K_{1,1+b_1,\ldots,1+b_n}(\tau)}
Indeed, as mentioned above, the odd size $2n+1$ link patterns $\pi'$ are obtained from those
of size one more $2n+2$ $\pi$ by removing the rightmost arch and leaving its beginning
point unmatched, so arch closings indeed get shifted by one unit in this bijection, namely $b_1(\pi)=1$,
and $b_i(\pi)=b_{i-1}(\pi')+1$ for $i=2,\ldots,n+1$. 
This makes the connection between Eqs.~\infiodd\
and \recast\ explicit.

The results \zerodd--\infiodd\ 
translate immediately into the corresponding identical small and large $\tau$ estimates for
$\Psi_\pi(\tau)$, as the change of basis from arch openings to link patterns is directly inherited from that of the even case $2n+2$.

\subsec{Generalized sum rule}
An analogous statement as that of Sect.~5.3 may be derived
for the case of odd size $2n+1$, in which we now have to sum over ``arch closing'' basis elements
$K'_{b_1,\ldots,b_n}$ with $b_i=2i-\epsilon_i$, $\epsilon_i\in\{0,1\}$.
The generalized sum rule now reads:
\eqn\considodd{K'(t \vert \tau)\equiv \sum_{\epsilon_i\in\{0,1\}} t^{\Sigma \epsilon_i} 
K'_{2-\epsilon_1,4-\epsilon_2,\ldots,2n-\epsilon_n}(\tau)}
and reproduces $\sum_{\pi\in LP_{2n+1}} \Psi_\pi$ at $t=1$.

Repeating exactly the same steps as those of Section 5.3, with now an extra factor of 
$\prod_{m=1}^n (1+\tau u_m+u_m^2)/u_m$, we easily arrive at:%
\eqnn\Kprime
$$\eqalignno{K'(t\vert \tau)&=\det_{1\leq \ell,m\leq n}  \left[ \oint {du \over 2\pi i u^2} 
(1+u^{-1}(\tau+u^{-1}))(1+t u)u^{\ell-m} (\tau+u^{-1})^{m-1}(\tau+u)^{\ell-1} \right]\cr
&=\det_{1\leq \ell,m\leq n}  \left( \phi_{\ell,m}(t\vert \tau)+\phi_{\ell,m+1}(t\vert \tau)\right) 
&\Kprime\cr}
$$
where%
\eqnn\defphi
$$\eqalignno{\phi_{\ell,m}&= \oint {du \over 2\pi i u^2} (1+t u)u^{\ell-m} (\tau+u^{-1})^{m-1}(\tau+u)^{\ell-1} \cr
&=\sum_r \tau^{2\ell+2m-2r-4} {\ell-1\choose r-\ell}\left[ \tau {m-1\choose r+1-m}+t {m-1\choose r+2-m}\right]&\defphi\cr}
$$

It is straightforward to show that%
\eqnn\match
$$\eqalignno{K'(t\vert \tau)&=\det_{1\leq \ell,m\leq n}  \left( g_{\ell,m}(t\vert \tau)+g_{\ell-1,m}(t\vert \tau)\right) \cr
&=\det_{1\leq \ell,m\leq n}  \left(g_{\ell,m}(t\vert \tau)\right) 
&\match\cr}
$$
where we have defined
\eqn\defg{g_{\ell,m}(t\vert \tau)=\sum_r \tau^{2\ell+2m-2r-1} {\ell \choose r-\ell}\left[ \tau { m-1\choose 2m-r}+ 
t {m-1\choose 2m-1-r} \right]  }
The proof of \match\ is obtained by performing a term-by-term identification of $\phi_{\ell,m}+\phi_{\ell,m+1}$ 
with $g_{\ell-1,m}+g_{\ell,m}$, and then identifying the determinant of these entries with 
$\det(g_{\ell,m}(t\vert \tau))$
by column manipulations of the latter matrix.
%using the fact that the matrix $G$ with entries $g_{\ell,m}(t\vert \tau)$ 
%has the same determinant as the matrix $G P$, where $P$ has entries 
%$P_{\ell,m}=\delta_{m,\ell}+\delta_{m,\ell+1}$, and is obviously of determinant $1$.

\fig{(a) The fundamental domain of a CSTCPP and its NILP description (blue and black paths)
and (b) the correspondence (heights of the box piles) to triangular arrays of integers (c).
The color code for weights here is: 
red=$\tau$, purple=$t$, and green=$x$.}{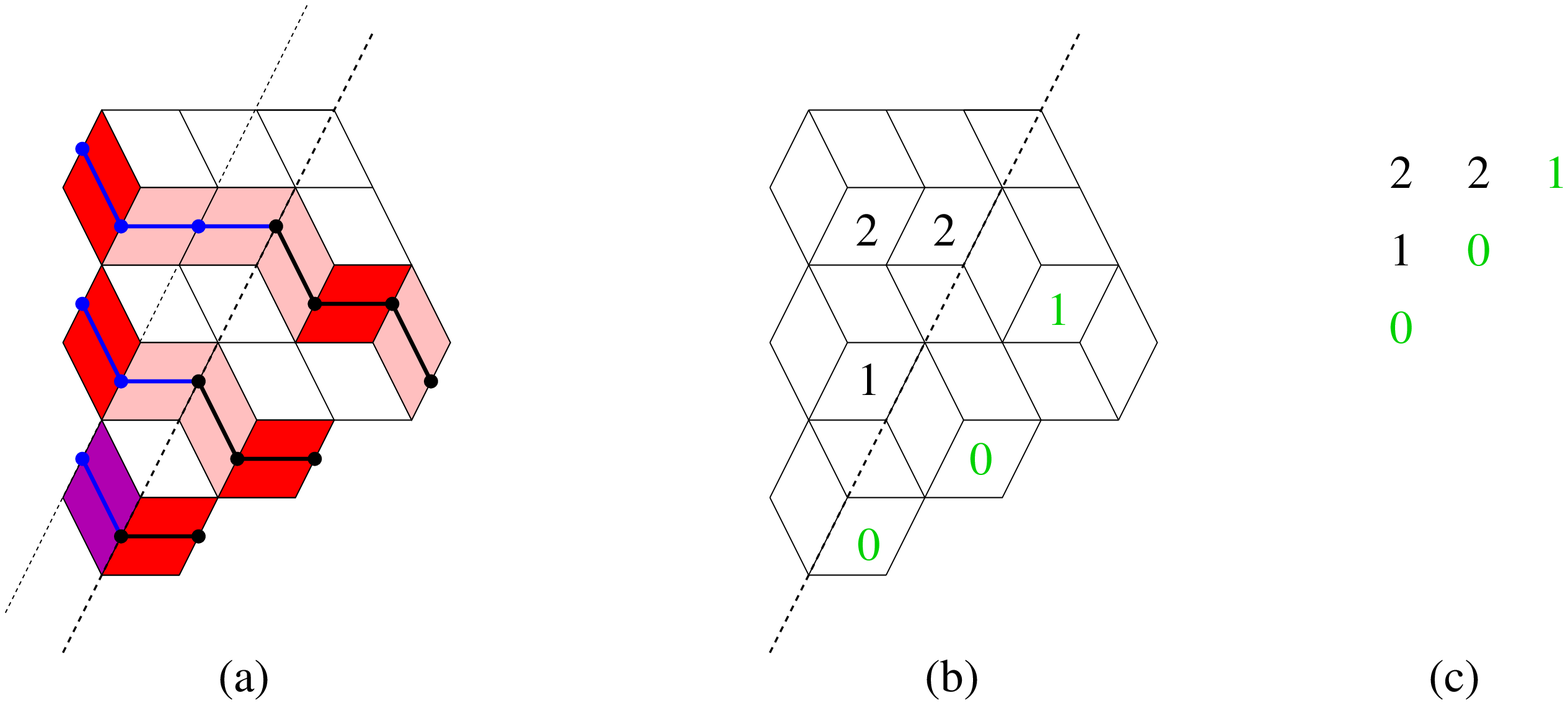}{10cm}\figlabel\cstcpparray
The NILP formulation of CSTCPP
given in Ref.~\DFVS\ is similar to the even case, except this time the paths are viewed as two
sets of paths of equal lengths (each in bijection with TSSCPP). 
The generating function $K'(t\vert \tau)$ corresponds to the generating polynomial of
these objects, with a weight $\tau$ for the same type of tiles as in the even case, see Fig.~\cstcpparray, 
{\it except}\/ in the last steps of one of the set of paths, where this weight is replaced by $t$.

The expression in the second line of \match\ for $t=1$ matches exactly the conjectured result of 
Ref.~\DFVS\ (Eq.~(4.2)), 
equal to the generating polynomial for weighted CSTCPP. 

Likewise, at $t=0$ one recovers the conjectured expression of Ref.~\DFVS\ (conjecture 4, Eq.~(5.3)), and
the limit $t\to \infty$ yields the corresponding reflected link pattern (conjecture 3, Eq.~(5.2)).
Note that the latter is identical to the sum rule $K(\tau^{-1}|\tau)$ in even size one less ($2n$) with
$t=\tau^{-1}$, due to simple identities relating $K'$ and $K$. Therefore it is also equal to
the generating function $T_n(\tau^2,1)$ of \Rob.

Finally at $t=\tau$ we can identify $K'(\tau|\tau)$ with
the generating function $T_n(\tau^2,0)$ defined in \Rob. Once again, this is no surprise since $T_n(x,0)$ is
the generating function for triangular arrays of non-negative integers
$a_{ij}$, $i,j\ge 1$, $i+j\le n$,
with weakly decreasing rows and columns and such that
$a_{i1}\le n-i$ for all $i$,
which turn out to be in one-to-one
correspondence with CSTCPP, see Fig.~\cstcpparray. As in the case of even size, one can check that the
refinements are the same, i.e.\ that
the weight $x$ given to lozenges of
one of the three types that are below the diagonal is the same weight that
is given to CSTCPP in $K'(\tau|\tau)$ if one sets $x=\tau^2$.
Note that once again simple identities
show that $K'(\tau|\tau)$ is equal to $K(0|\tau)$, that is the component $\Psi_{\pi_{max}}(\tau)$ of size $2n$.

\newsec{Conclusion}
In this paper, we have proved various conjectures regarding the minimal polynomial
solution of the $q$KZ equation with reflecting boundaries with $q$ generic, 
expressed in the link pattern basis. This was done by
writing the solutions as multiple residue integrals, modulo a triangular change of basis.
As both integrals and the change of basis are completely explicit, we therefore end up with
an explicit formula for each component $\Psi_\pi(z_1,...,z_N)$. We hope these expressions
will help us address the full Razumov--Stroganov conjecture which gives a conjectural
interpretation for each $\Psi_\pi$ in the homogeneous case $z_1=...=z_N=1$
(and $q=-e^{i\pi/3}$), and hopefully come up with a more general combinatorial 
interpretation of the polynomials $\Psi_\pi(\tau)$ in the homogeneous case for generic $q$.
Note that we have now a numerical recipe for calculating the $\Psi_\pi(\tau)$, via 
the explicit inversion of the change of basis, and an explicit generation of the integrals.

As stressed and proved in this paper, the above change of basis is independent of the 
details of the boundary 
conditions imposed in addition to the main exchange relation $t_i\Psi=(e_i-\tau)\Psi$.
These details are
simply reflected by the insertion of some specific symmetric function $F$ in the definition
of the integrals. The techniques of the present paper may therefore presumably be adapted
to include the other boundary conditions considered in \ORBI, parametrized by the root systems
of classical Lie algebras.

As shown in \DFZJc, generalizations of the Razumov--Stroganov sum rule have been
obtained and proved in the case of the level 1 $q$KZ equation pertaining to higher rank 
($sl_k$) algebras at specific values of the parameter $q$ ($q=-e^{i\pi\over k+1}$ and 
$q=-1$). We believe the construction of the present paper may be generalized to these cases,
and may lead to new combinatorial interpretations, such as generalizations of Plane Partitions.

Another direction of future research is to revisit the model of
{\it crossing}\/ loops considered in \refs{\DFZJb,\KZJ,\KZJb}, 
which is based on the Brauer algebra, and obtain
integral formulae in the same spirit as those of the present work. It would then be particularly
interesting
to understand their interrelation with the geometry of the Brauer loop scheme.

\centerline{\bf Acknowledgments}
We acknowledge the support
of European Marie Curie Research Training Networks ``ENIGMA'' MRT-CT-2004-5652, 
``ENRAGE'' MRTN-CT-2004-005616, ESF program ``MISGAM''
and of ANR program ``GIMP'' ANR-05-BLAN-0029-01. We wish to thank
J.~de~Gier and P.~Pyatov for pointing out the appearance in our work
of the generating functions of \Rob, as well as for sharing their current research with us.
We also thank R.~Kedem and C.~Krattenthaler for useful discussions.

\listrefs

\end